\documentclass[10pt,journal,compsoc]{IEEEtran}
\usepackage[utf8]{inputenc}
\usepackage{amsmath} % align environment
\usepackage{amssymb} % mathbb
\usepackage[boxed]{algorithm}
\usepackage[noend]{algpseudocode}
\usepackage{graphicx}
\usepackage{soul}
\usepackage{hyperref}
\usepackage{color}
\usepackage{comment}
\usepackage{amsthm}
\usepackage{caption}
\usepackage{subfig}
\usepackage{booktabs}
\usepackage{multirow}
\usepackage{paralist}
\usepackage{wrapfig}
\usepackage{listings}
\definecolor{deepblue}{rgb}{0,0,0.5}
\definecolor{deepred}{rgb}{0.6,0,0}
\definecolor{deepgreen}{rgb}{0,0.5,0}
\newtheorem{definition}{Definition}

\algnewcommand{\IfThenElse}[3]{% \IfThenElse{<if>}{<then>}{<else>}
	\State \algorithmicif\ #1\ \algorithmicthen\ #2\ \algorithmicelse\ #3}
\algnewcommand{\IfThen}[2]{% \IfThen{<if>}{<then>}
	\State \algorithmicif\ #1\ \algorithmicthen\ #2}
\algnewcommand{\ElseThen}[1]{% \ElseThen{<if>}{<then>}
	\State \algorithmicelse\ #1}

\begin{document}
	\title{PySchedCL: Leveraging Concurrency in Heterogeneous Data-Parallel Systems}
	\author{Anirban Ghose, Siddharth Singh, Vivek Kulaharia, Lokesh Dokara, Srijeeta Maity and Soumyajit Dey
	\IEEEcompsocitemizethanks{
	\IEEEcompsocthanksitem A. Ghose, S. Maity and S. Dey are with the Dept. of Computer Science and Engg., IIT Kharagpur Email: \{anirban.ghose, srijeeta, soumya\}@cse.iitkgp.ac.in
	\IEEEcompsocthanksitem S. Singh is with the Dept. of Computer Science, University of Maryland, College Park  Email: ssingh37@umd.edu
	\IEEEcompsocthanksitem V. Kulaharia is with Cohesity Solutions India Pvt Ltd., working as a Member of Technical Staff Email: Vivek.Kulaharia@cohesity.com
	\IEEEcompsocthanksitem L. Dokara is with iB Hubs working as a DevOps Specialist  Email: lokesh@ibhubs.co
	}
    
	}

% 	\author{Anirban Ghose}
% 	\author{Siddharth Singh}
% 	\author{Vivek Kulaharia}
% 	\author{Lokesh Dokara}
% 	\author{Srijeeta Maity}
% 	\author{Soumyajit Dey}
	\maketitle         
	
\begin{abstract} In the past decade, high performance  compute capabilities exhibited by heterogeneous GPGPU platforms have led to the popularity of  data parallel  programming languages such as CUDA and OpenCL. Such languages, however, involve a steep learning curve as well as developing an extensive understanding of the underlying architecture of the compute devices in heterogeneous platforms. This has led to the emergence of several High Performance Computing frameworks which provide high-level abstractions for easing the development of data-parallel applications on heterogeneous platforms. However, the scheduling decisions undertaken by such frameworks only exploit coarse-grained concurrency in data parallel applications. In this paper, we  propose {\em PySchedCL}, a framework which explores fine-grained concurrency aware scheduling decisions that harness the power of heterogeneous CPU/GPU architectures efficiently. %, a feature which  is not provided by existing HPC frameworks. 
We showcase the efficacy of such scheduling mechanisms over existing coarse-grained dynamic scheduling schemes by conducting extensive experimental evaluations for a Machine Learning based inferencing application. 
	\end{abstract}
	
	% \begin{keyword}
	% Heterogeneous Multi-core architectures \sep  Static Analysis \sep Machine Learning \sep OpenCL \sep Scheduling.
	% \end{keyword}
	
\section{Introduction} %2.5 pages
The rise of data parallel programming languages like OpenCL \cite{stone2010opencl} and CUDA \cite{nvidia} have paved the way for high throughput application development on  big data clusters as well as embedded platforms comprising multiple CPU and GPU cores. Such  frameworks support asynchronous event driven programming models that enable both data parallel and task parallel paradigms of computation for implementing high performance parallel applications. %\sd{The data parallel programming model has provisions for implementing a computational kernel which represents the core computation for a given algorithm. A data parallel computational kernel launches multiple threads in parallel across multiple SIMD enabled compute units. Each thread applies the specified kernel  transformation to designated data points of the input data space. The task parallel programming model supports parallelism at the task/kernel level where application task graphs comprising multiple kernels with dependencies, each representing distinct computational transformations can be dispatched and executed on  multiple devices in a target heterogeneous platform. -GYAN ??? UNNECESSARY?} 
The OpenCL runtime system additionally has provision for program portability across different types of devices i.e. the same computational kernel source code can be compiled into device specific binaries for execution on different devices. 
	\par
\begin{figure}[ht] 
		\centering
		\includegraphics[scale=0.35]{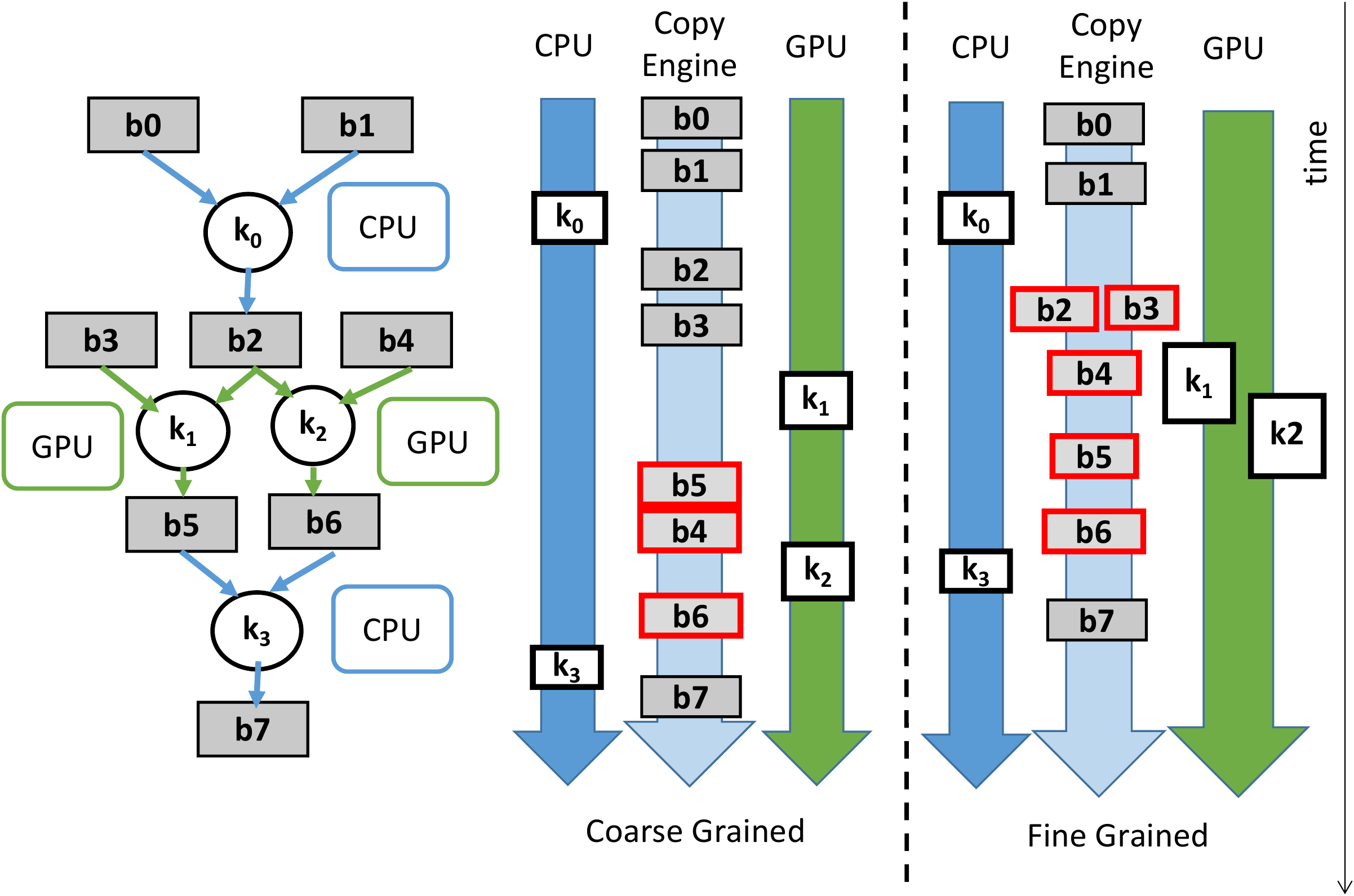}
		\vspace{-2mm}
		\caption{DAG Mapping Decisions \label{fig:dagmapping}}
	\end{figure}

 Given heterogeneous platforms comprising multiple devices of varying computational power, determining efficient architecture-to-application mapping decisions require extensive domain knowledge of platform level characteristics as well as precedence constraints enforced by the application which is typically represented as a directed acyclic graph (DAG) of tasks.  As an illustrative example, let us consider a simple fork-join DAG in Fig. \ref{fig:dagmapping}.
	\par We consider a heterogeneous platform comprising a single CPU and a single GPU along with a DMA copy engine responsible for transferring data across the PCI-Express bus from the CPU to the GPU and back. The fork-join graph comprises four tasks, each representing some computational kernel which takes as input two input buffers and produces one output buffer. In Fig. \ref{fig:dagmapping}, the rectangular nodes represent input and output buffers and the circular nodes represent kernels. We use this convention throughout the paper. The edges between a buffer and task represents the precedence constraints between tasks as well. Given a heterogeneous compute platform comprising a single CPU and a single GPU there can exist a total of 16 task-device mappings for this DAG where task(s) are  either mapped to a GPU device or a CPU device. In Fig. \ref{fig:dagmapping}, we explore one of the 16 possible mappings where $k_0$ and $k_3$ are  mapped to a CPU device, $k_1$ and $k_2$ are mapped to a GPU device. 
    	\par Scheduling decisions for general application DAGs are coarse-grained in the sense that  each task is mapped to a single device at a time and the associated kernel execution, buffer reads and writes are finished completely before proceeding to execute successors of the kernels. In Fig. \ref{fig:dagmapping}, for kernel $k_2$ to start execution, $k_1$ must finish and the copy engine should copy the resultant buffer $b_5$ to the host. After that the required input buffer $b_4$ has to be copied to the GPU device. The scheduling decisions are achieved by designing complex host programs that orchestrate the process of mapping individual kernels to target devices of the heterogeneous platform while maintaining precedence constraints. Alternatively, there exist several frameworks proposed in the recent past that alleviate the burden of implementing such complex orchestrators for undertaking coarse-grained scheduling decisions. The frameworks can be classified into two broad categories.  The first category of frameworks  \cite{pekka,henry2014toward} provide a top-level API as well as additional programming constructs using which a designer can explicitly specify the mapping and scheduling of OpenCL kernels without writing a complex host program. The second  category of frameworks (StarPU, MultiCL) \cite{augonnet2011starpu,multicl} provide  scheduling engines optimized for heterogeneous clusters with support for custom scheduling heuristics. These frameworks require as input the DAG specification and a scheduling heuristic for mapping OpenCL kernels optimally on a heterogeneous platform. Both styles rely on deriving coarse-grained scheduling decisions for application DAGs.  
	\par In contrast, we believe scheduling decisions should be more fine-grained in nature allowing execution of multiple tasks in the same device and interleaving copy operations with execute  operations. This is exemplified in the right hand side scheduling option of Fig. \ref{fig:dagmapping}. In the figure we show the execution timings of kernels in CPU and GPU  along with the data  transfers scheduled in the available copy engine in the GPU platform.	We can observe for kernel $k_1$, the two input buffers $b_2$ and $b_3$ can be transferred by the copy engine in parallel. Also while kernel $k_1$ is executed,  $b_4$ can be transferred asynchronously to the GPU device. The kernel $k_2$ executes in parallel with $k_1$ while sharing the same GPU resource. As a result, we observe that the individual times of $k_1$ and $k_2$ increase. However, the overall time to finish DAG execution decreases. 
	\par Implementing such fine-grained scheduling requires designing an even more complex host program capable of  i) asynchronously interleaving data transfers as and when required and ii) clustering multiple tasks to the same device as and when feasible. These scheduling decisions can be achieved by setting up multiple worker queues per device and asynchronously enqueueing commands for executing multiple kernels on the same device. For the CUDA runtime system, these worker queues are referred as {\em CUDA streams}.  For the OpenCL runtime these are referred as {\em command queues}. Naturally, the end user has to consider the computational capability of the device and the individual computational requirements of each concurrent kernel before dispatch. 
	\par We propose {\em PySchedCL},  a platform agnostic programming framework which is possibly the first computer-aided design solution that is capable of automating the process of deriving both coarse-grained and fine-grained scheduling decisions for efficient collaborative execution of application DAGs on heterogeneous multicores comprising CPU and GPU devices. The proposed framework supports reduction of considerable implementation overhead and automatically outputs scheduling decisions that exploit concurrency, requiring minimal intervention from the programmer. The framework is built using the widely used PyOpenCL API \cite{pyopencl} and facilitates rapid development and deployment of OpenCL applications. We choose OpenCL, since it offers device portability, thus supporting a myriad of compute devices such as CPU, GPUs, FPGAs, DSPs etc.  Our framework enables the user to concentrate only on developing OpenCL kernels and perform minimum manual intervention that would help in finally determining near optimal runtime scheduling decisions for data parallel applications on a target heterogeneous CPU-GPU platform. We note the optimizations proposed are generic and the ideas can be leveraged for any data parallel heterogeneous setting. The salient features of the proposed framework are enumerated as follows.   
	
	\begin{compactenum}
		\item %The framework supports a design frontend that  facilitates programmers to develop and execute application task graphs without having to consider the intricacies of runtime environment. 
		The framework eases development of data parallel applications through specification files, thus completely bypassing the requirement of manually implementing orchestrator host programs.  % The task of the application designer is only to develop individual kernels and populate these specification files. 
		As a representative example, one can implement the host code of a Transformer Neural Network based inference pipeline \cite{DBLP:journals/corr/VaswaniSPUJGKP17} which takes $\approx 130$  lines of vanilla OpenCL code using a specification file of $25$ lines in our framework. 
		%	\item The framework supports a feature analysis module coupled with a machine learning module inspired from existing works for statically determining optimal application-to-architecture mapping decisions to be used at runtime.
		\item The framework comprises a customizable scheduling backend optimized for automatically extracting fine-grained concurrency in applications executing on heterogeneous compute platforms. The backend also enables programmable scheduling with its rich API support, thereby allowing users to design, experiment and validate both coarse-grained and fine-grained scheduling policies on top of the default strategies in the framework .
%executing an application comprising either a single kernel or multiple kernels with dependencies efficiently.  %The scheduling engine  mimics the behaviour of the orchestrating host program and has support for clustering kernels in a task graph as well as automatically making decisions regarding concurrent kernel execution. 
%\item Apart from the default schemes of the scheduling engine, the framework also provides a set of API functions that can be overriden by%with the framework seamlessly taking care of the intricacies of runtime execution. %automatically sets up for device worker queues   for exploiting application level fine-grained concurrency on heterogeneous compute targets. 
\item While the usability of the framework is evident in terms of programming effort, we also observe the efficacy of its automated fine-grained scheduling schemes through extensive experimentation for AI workloads such as transformer networks. We have observed considerable speedups in the range of $1.4-3.4\times$ for such applications compared to existing coarse-grained scheduling approaches supported in frameworks such as StarPU \cite{augonnet2011starpu}, SOCL  \cite{henry2014toward} etc. %which employ standard  list scheduling approaches. 

		%\item In addition to the modules above, the framework presents a generic API for programmers to implement and  experiment with a plethora of scheduling policies and ascertain which policies would prove to be beneficial for executing general purpose OpenCL applications on heterogeneous CPU-GPU platforms while optimizing for various scheduling metrics such as maximizing throughput, minimizing schedule makespan etc. We propose relevant heuristics that exploit locality of device memory spaces as well as concurrent kernel execution and provide extensive experimental results for the same. 
	\end{compactenum}  
	
	%The framework thus eases complex OpenCL application  development with the help of specification files and performs fine/coarse-grained mapping and scheduling of data-parallel kernels on OpenCL compliant devices in  heterogeneous CPU-GPU platforms. 
The remainder of the paper is organized as follows. In Section \ref{sec:OpenCL}, we  present necessary background on OpenCL runtime and the inadequacy of existing OpenCL based high-level scheduling frameworks, thus motivating the requirement of our proposed framework. This is followed by problem formulation in Section \ref{sec:prob} and  the software architecture of the framework in Section \ref{sec:framework}. We perform extensive experimentation and provide a comparative evaluation in Section \ref{sec:expt}. We present a comprehensive list of related work in Section \ref{sec:rel} and finally conclude the paper in Section \ref{sec:conclusion}. 
	
\section{Background and Motivation} \label{sec:OpenCL} %4 pages
Any OpenCL application typically comprises two distinct program entities - i) the {\em host} which is a single threaded sequential program executing on one CPU core that orchestrates the entire process of managing data and issuing directives for parallel execution, and ii) kernel(s) which execute on devices with support for vector processing (CPU,GPU,FPGA,DSP etc). For every computational kernel, the single-threaded host program leverages command queues supported by the OpenCL API to issue commands for  performing the following operations - i) copying the data from host to input buffers resident on device memory (Host to Device or H2D transfer), ii) launching multiple instances of the same kernel to process the data copied to the device and iii) copying back the data stored in output buffers in the device back to the host memory after the kernel has finished processing  (Device to Host or D2H transfer). 
	\par
\begin{figure}[ht]  
	\centering
	\includegraphics[scale=0.47]{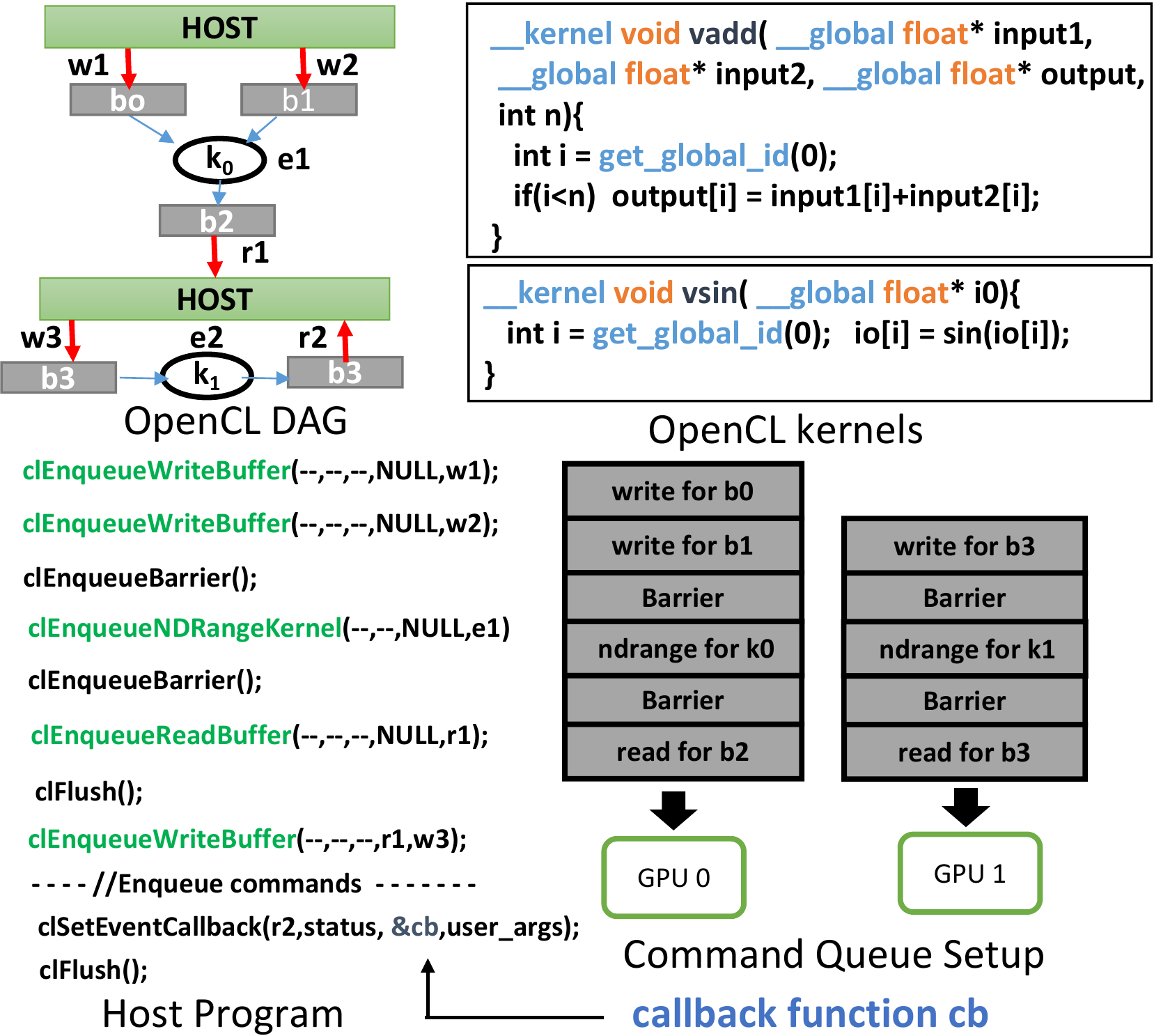}
	\vspace{-2mm}
	\caption{OpenCL Execution \label{fig:OpenCLArch}}
\end{figure}

 As an illustrative example, we consider a simple OpenCL application DAG depicted in the top left of Fig. \ref{fig:OpenCLArch} where kernel $k_0$ performs a vector addition operation ($vadd$) and kernel $k_1$ performs a simple element-wise trigonometric sine operation ($vsin$) on the output of $k_0$. The kernel $k_0$ takes as input two input buffers ($b0$ and $b1$), performs element-wise addition and produces an output buffer ($b2$). The kernel $k_1$  takes one buffer ($b3$) and performs an inplace element-wise sine operation. The corresponding kernel codes of  $k_0$ and $k_1$ are depicted in the functions $vadd$ and $vsin$ respectively in the top right subfigure of Fig. \ref{fig:OpenCLArch}. 
\par Considering a heterogeneous platform comprising two GPU devices $GPU_0$ and $GPU_1$, it may be observed in the bottom right subfigure of Fig. \ref{fig:OpenCLArch} that   two command queues are setup, one for each GPU device. Kernel $k_0$ executes on $GPU_0$ and kernel $k_1$ executes on $GPU_1$. Each command queue consists of a sequence of commands for H2D transfers, kernel execution and D2H transfers pertaining to each kernel. The associated host program  depicted in the bottom left of Fig. \ref{fig:OpenCLArch} illustrates how the command queues are populated. 
\par For $GPU_0$, the host first issues two {\em write} commands ({\tt clEnqueueWrite()}) for copying data from the host to the buffers $b0$ and $b1$. This is followed by a {\em barrier} directive ({\tt clEnqueueBarrier()}). The {\em barrier} command in general ensures that all commands enqueued previously finish before proceeding to execute commands enqueued after the barrier. In this case, it is ensured that the write commands are finished before processing the next command in the queue. The host next enqueues a kernel execution command or {\em ndrange} command using the function {\tt clEnqueueNDRangeKernel()} which spawns a collection of threads referred as \textit{work items} where each work item executes the function bodies depicted in the top right of Fig. \ref{fig:OpenCLArch}. The execution command is followed by a barrier directive and finally one read command ({\tt clEnqueueReadBuffer()}). In a similar fashion, for the command queue of device $GPU_1$, the host issues a write command (for buffer $b3$), an execute command for $vsin$ kernel and  read command (for buffer $b3$). We note that barriers in general incur synchronization overhead. For the remainder of the paper we assume that barriers are not enqueued and that command queues follow inorder execution i.e. commands are executed in the order which they are enqueued and each command cannot start until the previous command has finished completely.   
%\vspace{-3mm}
%\vspace{-2mm}
%on a \textit{processing element} on the heterogeneous platform. %Each work item is referred by a unique identifier $i$ obtained using the {\tt get\_global\_id()} OpenCL function and is responsible for the addition of  data points in the two input buffers $b0$ and $b1$ ({\tt input1[i]} and {\tt input2[i]} in function call) and storing the result in the corresponding location of the output buffer ({\tt output[i]} in function call). %Work items are further grouped into \textit{work groups} and each work group is scheduled for execution on a \textit{compute unit} in an OpenCL compliant device. A \textit{compute unit} may be a Symmetric Multiprocessor(SM) for a GPU device, a single core of a multicore CPU etc. 
 
\par The OpenCL runtime system uses {\em event}  objects for enforcing dependencies across multiple commands resident in same or different queues. In Fig.  \ref{fig:OpenCLArch}, the $i^{th}$ \textit{write} for the H2D transfers (highlighted by red edges in the OpenCL DAG) are associated with event $w_i$. In a similar fashion, the $i^{th}$ \textit{ndrange} command for kernels and $i^{th}$ \textit{read} command for D2H transfers (highlighted by red edges in the OpenCL DAG) are associated with events $e_i$ and $r_i$ respectively. These are labelled in the OpenCL DAG. The associated event for a command is specified in the last argument of a function call  with the {\tt clEnqueue} prefix in the host program. The second last argument of each such function represents events on which the event associated with command $c$ is dependent for execution. For our representative example, from the OpenCL DAG it may be observed that the event $w3$ is dependent on $r1$, i.e. the write command corresponding to $w3$ should take place only when the read command corresponding to $r1$ is complete. This is specified in the {\tt clEnqueueWriteBuffer} command for buffer $w3$. 
\par Finally, for the event $r2$ associated with the last read command (`read for b3'), a \textit{callback function} $cb$ is setup using the {\tt clSetEventCallback()} API call. 
A callback function $cb$ in general is registered with any event $ev$. The object to the event along with data to be accessed by the callback function once it starts execution is specified in  $user\_args$. The callback function is spawned on a separate thread in parallel with the host program once the event $ev$ (denoting computation/data transfer) has achieved some status (any one of $enqueued$\,/\,$submitted$\,/\,$complete$) as specified in the argument $status$ and is implemented typically to notify the host. 
%\sd{It starts execution on a separate thread in parallel with the host program once the associated event $r2$ has achieved $complete$ execution status. In general, $cb$ is implemented to notify the host that some event (denoting computation/data transfer) has achieved some status  (any one of $enqueued$\,/\,$submitted$\,/\,$complete$) as specified in the argument $status$. Data specified in the $user\_args$ are accessible to the callback function once it starts execution. Typically, the event object is specified here.} 
In Fig. \ref{fig:OpenCLArch}, $cb$ notifies the host, once $r2$ has achieved $complete$ status i.e. `read for b3' has finished. We note the {\tt clEnqueue} OpenCL functions  enqueue operations  to each command queue for the devices,  i.e. after enqueuing the commands, the host is free to execute something else while those commands are executed on the target device.  %The state of the art frameworks (StarPU, MultiCL and SOCL) automate the process of setting up command queues and enqueuing commands for computational kernels allowing users to bypass the overhead of implementing boilerplate host code. However, the scheduling heuristics supported by these frameworks are optimized for GPGPU clusters comprising multiple compute nodes with heterogeneous CPU and GPU devices. The mapping decisions offered by these scheduling heuristics typically are coarse-grained in the sense that a single kernel can be mapped to a single device at a time. The supported heuristics do not leverage heuristics for concurrently executing multiple kernels on the same device. This can be achieved by setting up multiple OpenCL command queues on the same device and enqueuing respective write, execute and read commands. 
	\subsection{Motivation}
\begin{figure}[ht]  
		\centering
		\includegraphics[scale=0.40]{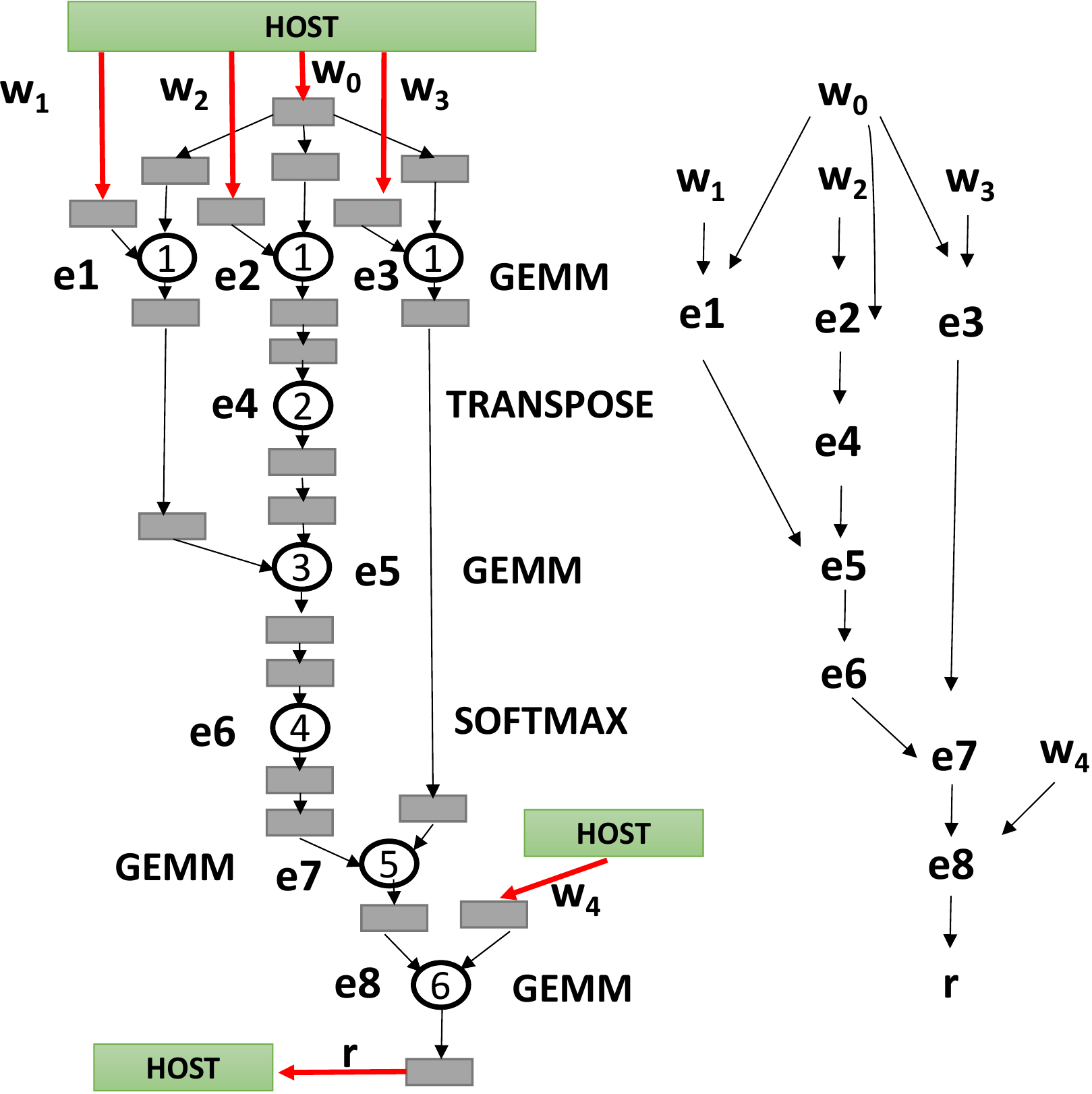}
		\vspace{-1mm}
		\caption{Event Dependencies for  DAG\label{fig:motivation0}}
	\end{figure}

	We consider a transformer application \cite{DBLP:journals/corr/VaswaniSPUJGKP17} which is a popular Deep Learning Neural Network pipeline for Natural Language Processing (NLP) tasks. The application exhibits ample scopes for exploiting concurrency with the possibility of executing multiple instances of standard General Matrix Multiply (GEMM) kernels in parallel. A sample DAG comprising 8 kernels for one layer of the transformer network is presented in Fig. \ref{fig:motivation0}. 
	\par  As per our earlier convention, the rectangular nodes represent input and output buffers and the circular nodes represent kernels. Each kernel is labeled with the corresponding level number starting from 1. Initially there is a copy operation which copies the same buffer to each of the kernels at level 1.  Each of the kernels in levels 1,4,5,6 represent General Matrix Multiply (GEMM) kernels where each kernel takes as input two buffers and produces one output buffer. The kernels in level 2 and level 3 represent transpose and softmax operations respectively, each processing one input buffer to produce one output buffer. The edges between rectangular nodes, i.e. buffers, represent data dependencies for the DAG. For enforcing precedence constraints between any pair of  kernels $(k_i,k_j)$, a programmer shall set event dependencies between read commands for output buffers of $k_i$ and write commands for input buffers of $k_j$, as was observed in Fig. \ref{fig:OpenCLArch}. For our transformer DAG depicted in Fig. \ref{fig:motivation0}, we assume that the entire DAG is mapped to a single GPU device. This implies that the output buffers of kernels at level $i$ to be processed as input buffers of kernels at level $i+1$, $i=[1,5]$ are already resident in GPU memory. Thus explicit reads and writes for dependent buffers between kernels in levels 1-5 are not required. In this scenario,  the programmer needs to set up event dependencies between \textit{ndrange} commands of kernels in levels $i$ and $i+1$ as depicted in the event dependency graph. One can observe from Fig. \ref{fig:motivation0}, that the actual H2D and D2H transfers (red edges) occur only for kernels at level $1$ and level $6$ while the remaining black edges reflect the input-output buffer dependencies between kernels in the DAG.
	\vspace{-2mm}
	\vspace{-4mm}
\par
\begin{figure}[ht]  
		\centering
		\includegraphics[trim=60 30 0 65 , clip, scale=0.45]{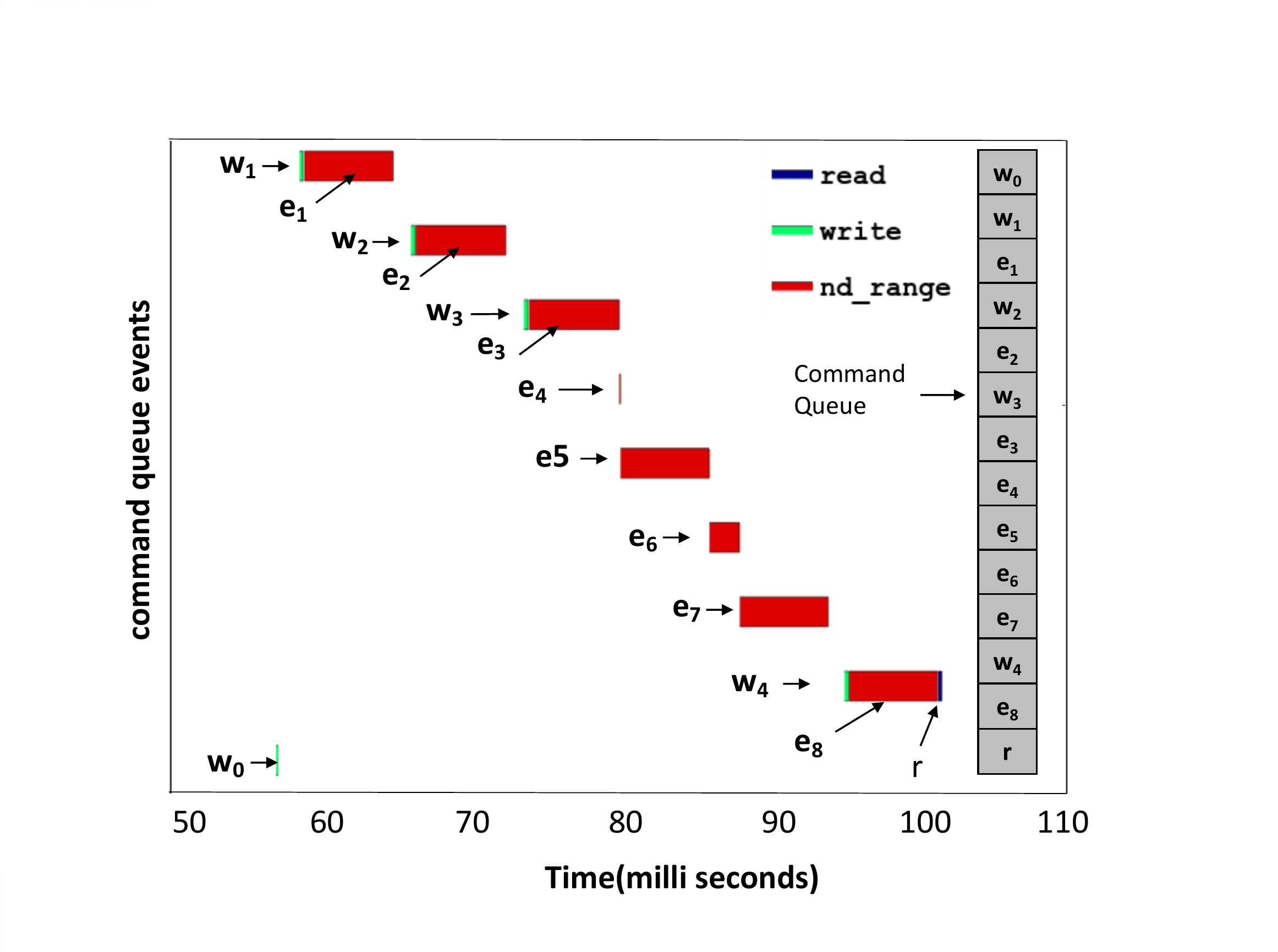}
		\vspace{-6mm}
		\caption{Coarse-grained Scheduling\label{fig:coarse}}
	\end{figure}

\begin{figure}[ht]  
		\centering
		\includegraphics[trim=120 30 0 10 , clip, scale=0.44]{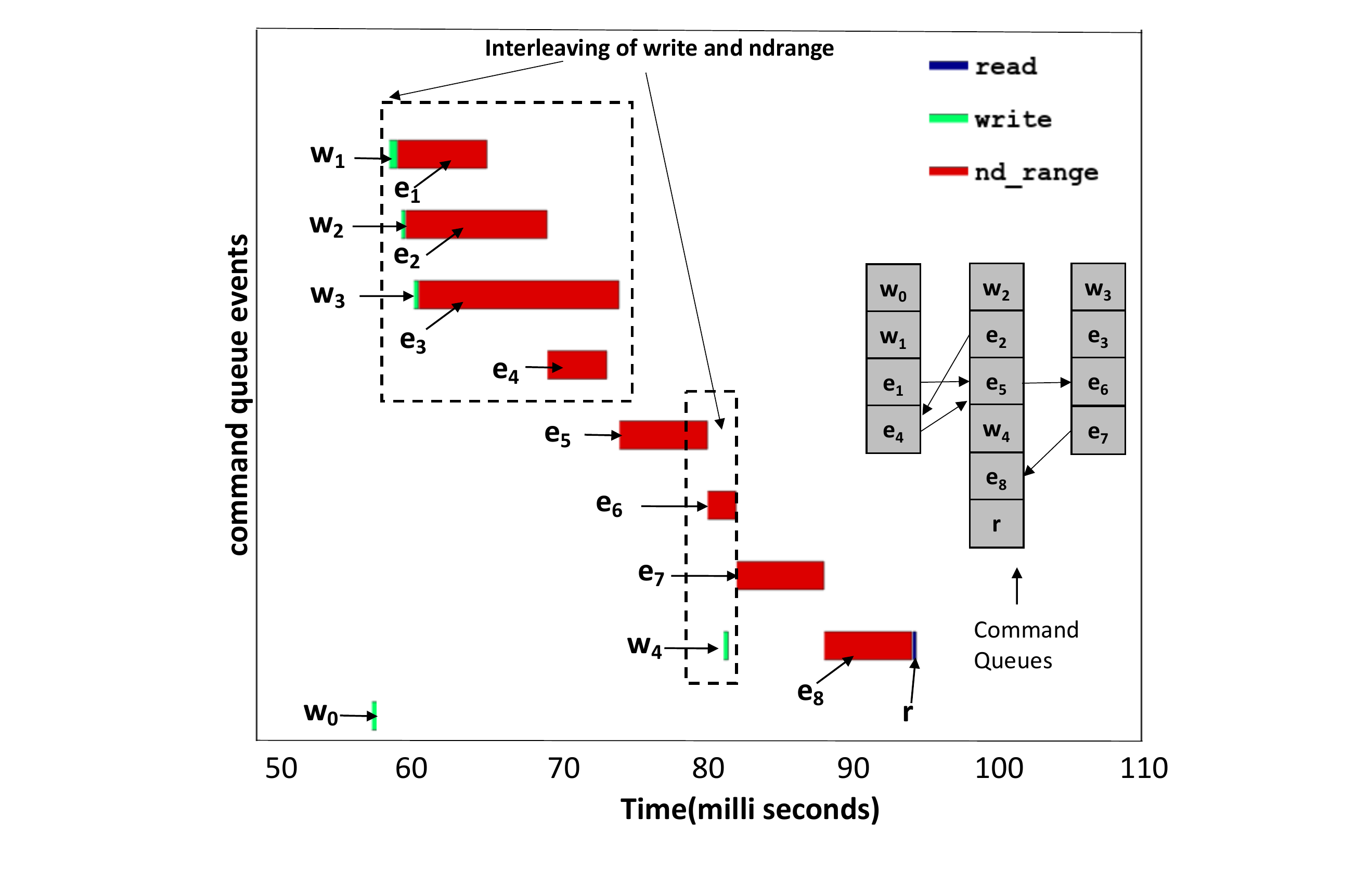}
		\vspace{-6mm}
		\caption{Fine-grained Scheduling\label{fig:fine}}
	\end{figure}

 In the left hand side of Fig 3. we label each kernel $k$ of the DAG with event $e_k$ associated with the corresponding \textit{ndrange} command for that kernel. Apart from this, we have a \textit{write} command $w_0$ responsible for copying one common buffer to be used for each GEMM kernel in level $1$. We also have \textit{write} commands $w_1,w_2,w_3$ for each of the remaining buffers required by GEMM kernels in level $1$ and a \textit{write} command $w_4$ for a buffer required by GEMM kernel in level $6$. Finally we have a \textit{read} command $r$ for the output buffer of the GEMM kernel in level $6$. The dependencies between these events are depicted in the corresponding event dependency graph in the right hand side of Fig. \ref{fig:motivation0}. The end designer is burdened with the task of manually writing a host program that will capture the event dependencies illustrated in this dependency graph for ensuring that precedence relations of the DAG are met during execution. This is achieved by using the complex programming constructs for OpenCL events and callback functions as discussed earlier.  Existing heterogeneous programming frameworks like StarPU \cite{augonnet2011starpu}, SOCL \cite{henry2014toward} and MultiCL \cite{multicl} all impose this same level of programming complexity. We next examine how coarse-grained and fine-grained scheduling decisions are made for mapping this DAG onto a single GPU device with the help of Figs. \ref{fig:coarse} and \ref{fig:fine} respectively. 
\par We execute the DAG on a heterogeneous platform comprising an NVIDIA GTX-970 GPU device and a Quadcore Intel i5-4690K CPU device.  Coarse-grained scheduling is achieved by setting up a single command  queue on the GPU device as depicted in the right hand side of Fig. \ref{fig:coarse}. As a consequence, all \textit{read}, \textit{write} and \textit{ndrange} commands for each of the 8 kernels as labelled by the events used in Fig. \ref{fig:motivation0} execute serially on the GPU device. In the left hand side of Fig.\ref{fig:coarse}, we plot a Gantt chart representing the time taken by the commands.  It is evident from the  Gantt chart that such serialized execution of commands on the GPU device result in an execution time of 105ms.
%\st{ We observe that the H2D transfer (associated with event $w0$) copies one buffer to the GPU device (highlighted by the red edge in Fig. \ref{fig:motivation0}) to be shared across all kernels in level $1$. The remaining H2D transfers occur for buffers of kernels in level $1$ and the kernel in level $6$. For GEMM kernels in level $1$, data is transferred for input buffers associated with events $w_1$, $w_2$ and $w_3$ and $w4$ respectively. Finally the last read associated with event $r$ occurs for the kernel in level $6$. REPETITION?}
\vspace{-4mm}
	\vspace{-6mm}
	\vspace{-2mm}
\par In contrast, if we set up multiple command queues, there is a possibility of leveraging fine-grained scheduling decisions that can i) interleave data transfers with \textit{ndrange} operations and can ii) execute multiple ndrange operations concurrently. In the right hand side of Fig. \ref{fig:fine},  we setup 3 command queues for achieving this. As a result of enqueuing the operations using multiple command queues, we observe from the corresponding Gantt chart in the left hand side of Fig. \ref{fig:fine},  several \textit{write} and \textit{ndrange} commands are interleaved thus resulting in an $8\%$ decrease in overall execution time with the DAG finishing in 95ms. On closer inspection, one can observe from the Gantt chart that while the \textit{ndrange} command associated with $e1$ is executing, the buffer associated with $w2$ can be copied simultaneously. This is because $w_2$ and $e_1$ belong to separate command queues in the right hand side of Fig. \ref{fig:fine}. In a similar vein, $w_3$ can also be copied while $e1$ and $e2$ are executing. Additionally, it can be seen that all kernels in level $1$ i.e. \textit{ndrange} commands associated with events $e_1$, $e_2$ and $e_3$ are  executing concurrently on the same device. We can observe something similar happening for the events $e_5$, $e_6$ and $w_4$ in the Gantt chart as well.  However, there are commands that despite belonging to different command queues are not  able to execute simultaneously due to the precedence relationships enforced by the event dependency graph illustrated in  of Fig. \ref{fig:motivation0}. These dependencies being part of the actual command queue setup, are represented by inter-queue edges between events in the command queue structure in the right hand side of Fig. \ref{fig:fine}. For example, since $e_4$ is dependent upon $e_2$, it can start execution only after $e_2$ has finished. But $e_4$ can still  overlap with $e_3$. Another interesting observation would be that the individual execution times for each kernel increases slightly as a result of interleaving. This is due to the fact, that different work groups of different kernels that have been concurrently dispatched  are scheduled in a round robin fashion to the compute units of the device, thus causing resource contention \cite{ccuda}. However, the total time for finishing kernels concurrently is lesser than the case when they are dispatched in sequence. We note that both the cases represented in Figs. \ref{fig:coarse} and  \ref{fig:fine} depict one of the possible command queue configurations and multiple such possibilities exist both for coarse-grained and fine-grained scheduling respectively.

\par The Gantt charts for the execution of DAG highlighted in Fig. \ref{fig:motivation0} thus reveal that fine-grained scheduling has potential performance benefits over coarse-grained scheduling in the context of data parallel programming models used for GPGPU systems. 

	\section{Problem Formulation}
	\label{sec:prob}
\begin{figure}[ht]
		\centering
		\includegraphics[scale=0.38]{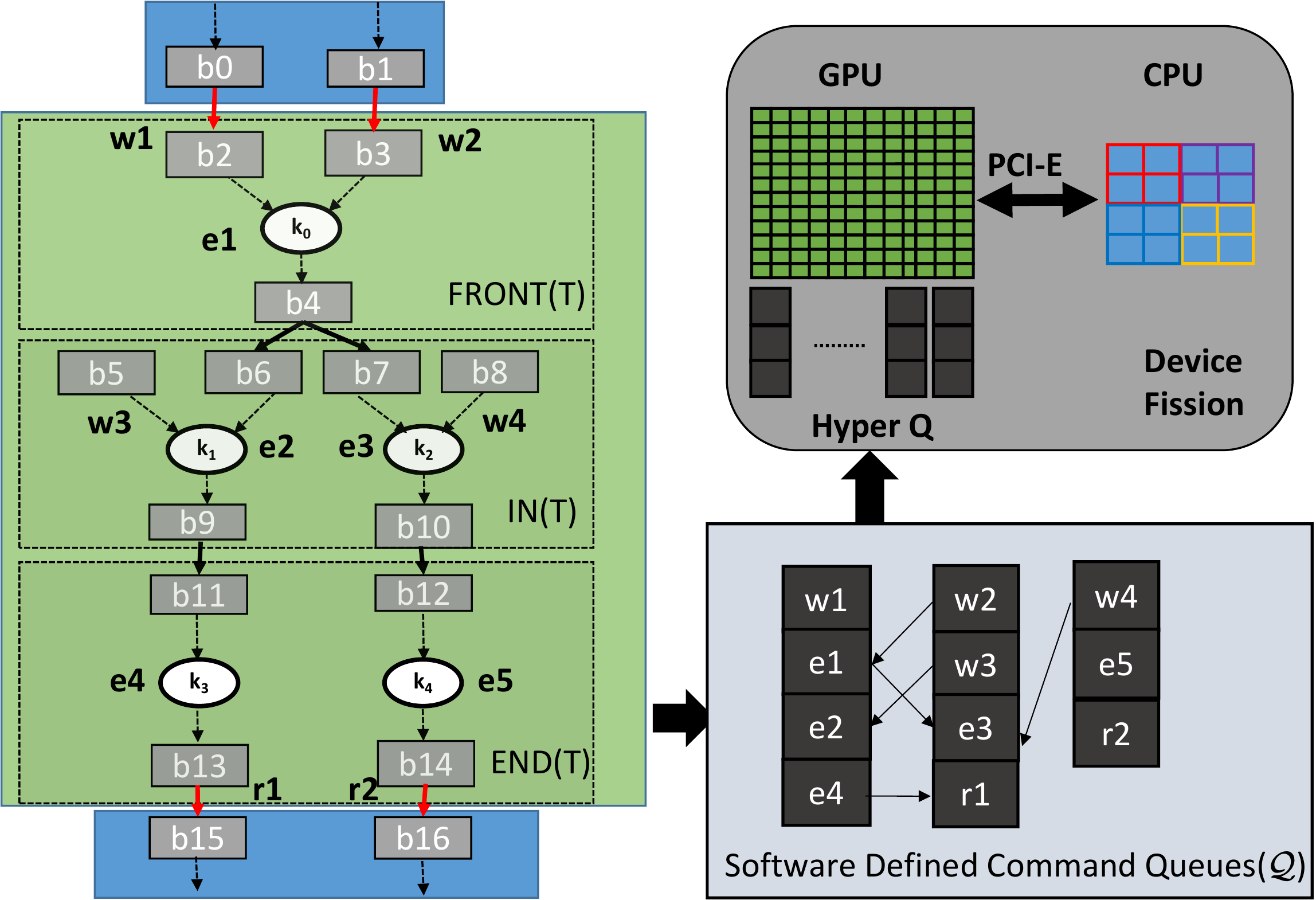}
		\vspace{-3mm}
		\caption{\small Platform and DAG Model\label{fig:platprob}}
	\end{figure}

	Let us consider a heterogeneous platform $\mathcal{P}$ depicted in Fig. \ref{fig:platprob} which comprises a CPU device and a GPU device connected via a PCI-Express bus. Each device has support for executing multiple kernels simultaneously. The OpenCL standard supports device fission for CPU devices i.e. a single CPU device can be partitioned into multiple subdevices, thereby enabling concurrent execution for the same. We consider as GPU an NVIDIA device with Hyper-Q support \cite{nvidia}. Hyper-Q offers a solution that allows the CPU host to dispatch multiple kernels simultaneously on the GPU device with the help of hardware managed work queues. 
	\par Let us represent an OpenCL application as a directed acyclic graph (DAG) $G = \langle (K,B),(E_I,E_O,E) \rangle$ where $K$ denotes the set of OpenCL kernels, $B= B_I \bigcup B_O$ represents the set of buffers for all $k \in K$. The set $B_I$ denotes the set  of input buffers and the set $B_O$ denotes the set of output buffers. The set $E_I \subseteq B_I \times K$ denotes the set of edge dependencies between each input buffer and kernel, $E_O \subseteq K \times B_O$ denotes the set of edge dependencies between each kernel and output buffer. The set $E \subseteq B_O \times B_I$ denotes the set of input output buffer dependencies across kernels in the DAG. For the remainder of the paper, we shall use this notation for representing DAGs.
	Command queues are typically setup per device, depending on which kernels are mapped to which devices. 
	\par Given an OpenCL DAG $G$, we denote a task component $T$ as a subset of kernels $K^\prime \subseteq K$  where each kernel $k$ is mapped to a device of the same type say $dev$. In our case, $dev= \{ cpu,gpu \}$. In Fig. \ref{fig:platprob},  $T  = \{k_0,k_1,k_2,k_3,k_4 \}$. For a given task component we define the following terminology.
	\begin{definition}
		Given a task component $T$ pertaining to some OpenCL DAG $G$, we define $FRONT(T)$ as the set of kernels where each kernel $k$ has input buffer dependencies $(b_i,k) \in E_I$ such that for $b_i$, if there exists an immediate predecessor $b_j$ where $(b_j,b_i)\in E$ and $(k^\prime,b_j) \in E_O$, then the kernel $k^\prime$ belongs to a different task component $T^\prime_{d'}$. 
	\end{definition}
	In Fig. \ref{fig:platprob}, we observe that $FRONT(T) = \{ k_0\}$, since both input buffers $b2$ and $b3$ have predecessors pertaining to kernels belonging in a different task component. 
	\begin{definition}
		Given a task component $T$ pertaining to some OpenCL DAG $G$, we define $END(T)$ as the set of kernels where each kernel $k$ has output buffer dependencies ($k,b_i) \in E_O$ such that for $b_i$ if there exists an immediate successor $b_j$ where $(b_i,b_j)\in E$ and $(b_j,k^\prime) \in E_I$ then kernel $k^\prime$ belongs to a different task component $T^\prime$.
	\end{definition}
	In Fig. \ref{fig:platprob}, we can observe that $END(T)=\{k_3,k_4\}$. \begin{definition}
		Given a task component $T$ pertaining to some OpenCL DAG $G$, we define $IN(T)$ as the set of kernels where each kernel $k \in T, k \notin FRONT(T), k \notin END(T)$.  
	\end{definition}
	In Fig. \ref{fig:platprob}, we can observe that $IN(T)=\{k_1, k_2 \}$. We classify buffer edge dependencies $(b_i,b_j) \in E$ into two categories -i) intra edge, ii) inter edge.
	Given a task component $T$ pertaining to a DAG $G$, an edge $(b_i,b_j)$ for kernels $k_i,k_j$ such that $(k_i,b_i) \in E_O$, $(b_j,k_j) \in E_I$ represents an intra edge if $k_i$ and $k_j$ belong to the same component and inter edge if they belong to different task components. In Fig. \ref{fig:platprob}, we can observe that $(b4,b6)$, $(b4,b7)$, $(b9,b11)$ and $(b10,b12)$ are intra edges, while  $(b0,b2)$, $(b1,b3)$, $(b13,b15)$ and $(b14,b16)$ are inter edges. 
	\par We classify kernel-buffer dependencies in $E_I$ and $E_O$ into two categories - i) isolated copy and ii) dependent copy. Given any kernel $k_i$, an edge ($b_i,k_i)\in E_I$ represents an isolated copy (write) {\em iff} for every $b_k \in B$, $(b_k,b_i) \notin E$. The same edge can represent a dependent copy if there exists some buffer $b_i \in B$ such that $(b_i,b_k)\in E$. In a similar fashion, an edge ($k_i,b_j)$ represents an isolated copy (read) {\em iff} for every $b_k \in B$, $(b_j,b_k) \notin E$ respectively and a dependent copy (read) if there existed some $b_k$ such that $(b_j,b_k) \in E$. In Fig. \ref{fig:platprob}, the edges $(b5,k_1)$ and $(b8,k_2)$ represent  isolated writes while every other kernel-buffer dependency represents dependent copies.
\begin{definition} Given a task component $T$ of an application DAG $G$ mapped to a device $d$ with $r$ command queues, we define the command queue data structure \textbf{$\mathcal{Q}$} = $\langle Q, E_q \rangle$  as follows.  $Q=\{q_1,q_2,\cdots,q_r\}$ is the set of command queues,  each command queue $q_i$ is a list such that every location $q_i[j] \in \{\textit{write},  \textit{ndrange}, \textit{read}\}$ contains any of these three commands  pertaining to some kernel belonging to $T$. Each element of $E_q$ is a precedence constraint of the form $\langle q_s[i], q_t[j] \rangle$,  $1\leq s \neq t \leq r$ which enforces that the $i$-th command enqueued in $q_s$ must finish execution before the $j$-th command enqueued in $q_t$ can start.
\end{definition}
\par A precedence constraint $\langle q_s[i],q_t[j] \rangle \in E_Q$ exists if any of the following is true - i) $q_s[i]$ is an isolated/dependent write $(b_l,k_m)$ and $q_t[j]$ is an ndrange operation for kernel $k_m$, ii) $q_s[i]$ is an ndrange operation for kernel $k_m$ and $q_t[j]$ is a dependent/isolated read $(k_m,b_l)$, iii) both $q_s[i]$ and $q_t[j]$ are ndrange operations for kernels $k_m$ and $k_n$ respectively such that there exists edges $(k_m,b_l) \in E_O$, $(b_p,k_n) \in E_I$  and $(b_l,b_p) \in E$ where $(b_l,b_p)$ is an intra edge. In Fig. \ref{fig:platprob}, $(b_3,k_0)$ corresponds to a dependent write for kernel $k_0$ thus requiring a dependency between associated operations $w_2$ and $e_1$ in \textbf{$\mathcal{Q}$}.  The edge $(e_1,e_3)$ represents the dependency between kernels $k_0$ and $k_2$ arising due to the dependencies $(k_0,b_4)$,$(b_4, b_7)$,$(b_7,k_2)$ where $(b_4,b_7)$ is an intra edge. The operations of a kernel $k_l \in T$ that are to be enqueued to some queue $q_s \in Q$ of \textbf{$\mathcal{Q}$} are determined by the framework using an enqueue procedure $enq(k_l,q_s)$ and is described as follows. 
\par \noindent i) If $k_l \in FRONT(T)$,  $enq(k_l,q_s)$ enqueues all dependent write commands for buffers $b_m$ corresponding to dependent writes $(b_m,k_l) \in E_I$ followed by ndrange command for $k_l$ to $q_s$.
\par \noindent ii) If $k_l \in END(T)$,  $enq(k_l,q_s)$ enqueues  ndrange command for $k_l$ followed by all dependent reads $(k_l,b_m) \in E_O$ for $b_m$ to $q_s$.   
\par \noindent iii) If $k_l \in IN(T)$,  $enq(k_l,q_s)$ only enqueues  ndrange for $k_l$ to $q_s$.
\par \noindent Note, a kernel may belong to any combination of the three sets discussed above. One can observe that since all kernels in $T$ are mapped to the same device, the $enq$ procedure using the rule set above ensures that redundant dependent reads from $FRONT(T)$, redundant dependent writes and reads from $IN(T)$ and redundant dependent writes from $END(T)$ are avoided from being enqueued. Apart from these enqueue operations, for every kernel $k_l$,  irrespective of which set it belongs to, $enq(k_l,q_s)$ enqueues to $q_s$ - (i) all isolated writes $(b_m,k_l) \in E_I$ for input buffers $b_m$ before enqueuing the ndrange command for $k_l$ and (ii) all isolated reads $(k_l,b_n) \in E_O$ for output buffers $b_n$ after enqueuing the ndrange command for $k_i$.
\par\noindent Note that the above rules for $enq$ when applied to individual kernels in a task component $T$ need not generate a unique command queue structure \textbf{$\mathcal{Q}$} for a device. This, coupled with different possible task component  partitions  based on device mapping decisions of the overall DAG lead to multiple possible dispatch orderings, i.e. scheduling decisions for the  kernels.
\begin{definition}
Consider a DAG $G = \langle (K,B),(E_I,E_O,E) \rangle$,  with a set of task components  $\mathcal{T}=\lbrace T_1, T_2, \cdots, T_M \rbrace$ such that $\bigcup_i T_i = K$, a heterogeneous CPU-GPU multicore target platform $\mathcal{P} = \{d_1,d_2,\cdots d_p \}$ containing $p$ devices, and the number of command queues for each device $\{r_1,r_2,\cdots r_p \}$ as given. Consider, the set of all command queues $\mathcal{R}= \{q_1,q_2,\cdots,q_N \}$ where $N=\sum_i r_i$. For such an application-architecture pair, a (valid) \textbf{schedule}  $\sigma$ is a collection of enqueue procedures $\{enq(k_i,q_j) | k_i \in K, q_j \in \mathcal{R} \}$ such that each kernel  $k_i \in K$ is dispatched in a topologically sorted fashion with respect to the ordering of $k_i$'s enforced by the edges in $G$.%,  $k_1 \preceq k_2 \preceq \cdots,k_{|K|}$.
\end{definition}
Our framework facilitates {\em automated} creation of such  correct-by-construction valid schedules, both coarse-grained and fine-grained.

%\par In the remaining sections of the paper, we discuss the intricate details of both the frontends and the backends of the framework. 

\section{Software Architecture} \label{sec:framework}
\begin{figure}[ht]  
		\centering
		\includegraphics[scale=0.35]{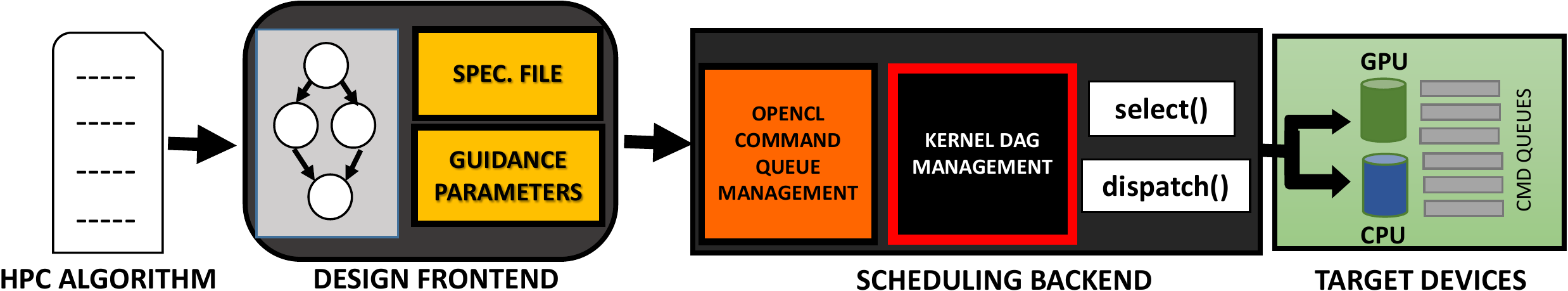}
		\vspace{-1mm}
		\caption{PySchedCL Toolflow\label{fig:pyschedcl}}
	\end{figure}

	An overview of the software architecture for {\em PySchedCL} is shown in Fig. \ref{fig:pyschedcl}. The framework comprises two distinct modules, the functionalities of which are elaborated below.
	\vspace{-2mm}
	\vspace{-5mm}
\par
\begin{figure}[ht]  
		\centering
		\includegraphics[scale=0.37]{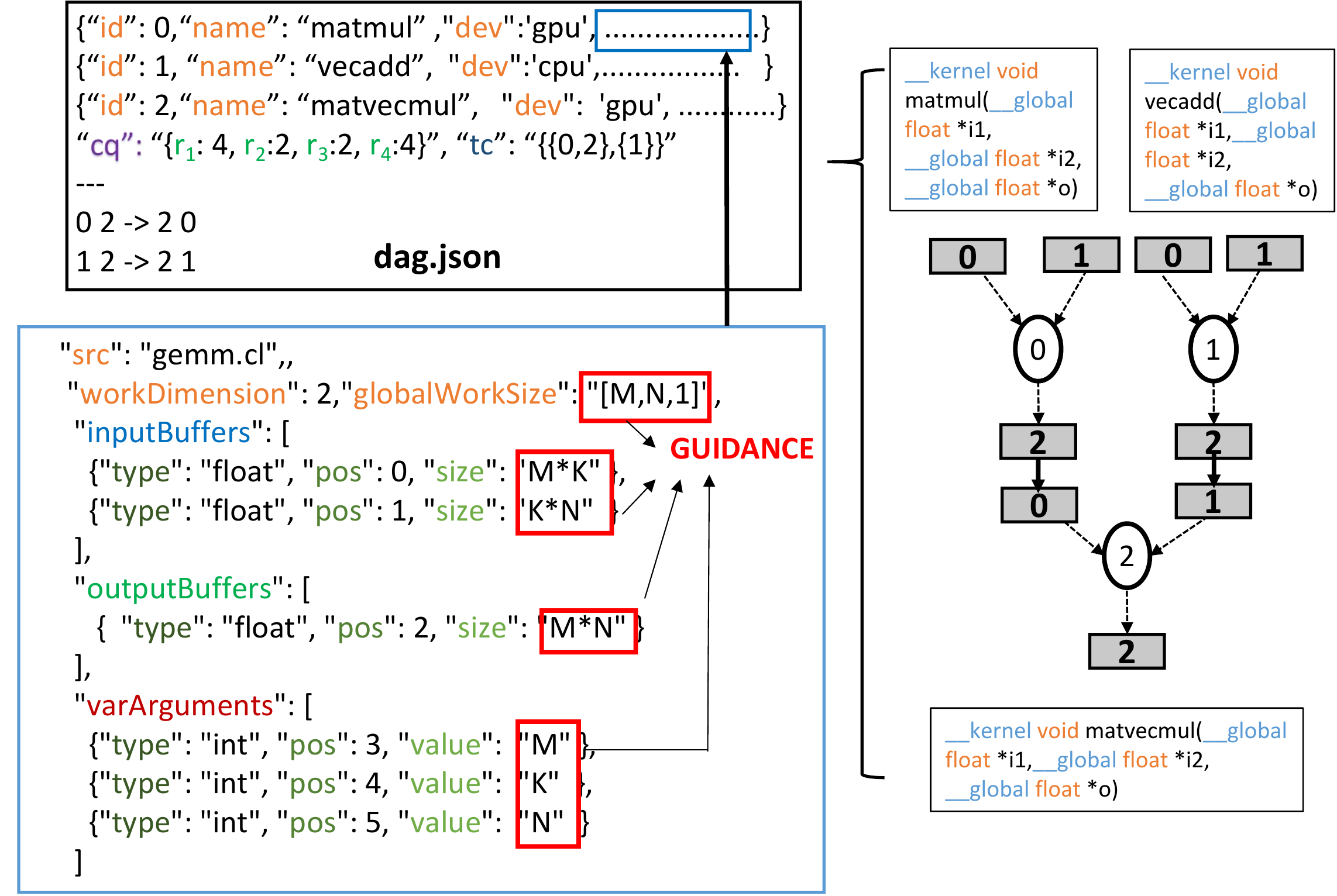}
		\vspace{-2mm}
		\caption{JSON Specification File for DAG \label{fig:json}}
	\end{figure}

 \noindent \textbf{A. Design Frontend: } The input to the scheduling framework is an OpenCL application represented in the form of an OpenCL DAG $G$ as discussed earlier. The proposed framework supports a specification file using which programmers can easily design an OpenCL application for execution on a heterogeneous platform. The specification file contains necessary buffer-kernel dependency information for an OpenCL DAG, along with necessary attribute information such as input/output buffers, variables passed as arguments for each constituent kernel. The file uses  Javascript Object Notation (JSON) format. Let us consider an example DAG comprising three kernels as depicted in Fig. \ref{fig:json}. Each kernel in {\tt dag.json} file as shown in top-left box of Fig. \ref{fig:json}, is designated  with i) a unique identifier field called $id$, ii) a $name$ file depicting the name of kernel function, iii) a device field $dev$ indicating the device type to which the kernel should be mapped (`cpu' or `gpu').  %The kernel with id $0$ represents matrix multiplication kernel which takes as input two matrices of dimensions $M \times K$ and $K \times N$ and produces an output of dimension $M \times N$. The kernel with id $1$ represents a vector addition kernel which takes two vectors of size $N$ and produces one output vector of size $N$. The kernel with id $2$ represents  matrix-matrix multiplication kernel which takes as input an $M\times N$ matrix and $N\times 1$ vector and produces an $M \times 1$ vector. The outputs of the kernels $1$ and $2$ are used by the kernel with id $2$. 
The task component partitioning $\mathcal{T}=\lbrace T_{1}, T_{2}, \cdots, T_{M}\rbrace$ for the DAG is specified as a list $tc=\{\{\cdots\},\cdots,\{\cdots\}\}$ with each sub-list $i$ being an enumeration of the kernel $id$-s in $T_{i}$. All kernels mapped to a task component must be given the same device type. In Fig. \ref{fig:json}, the list $tc =\{\{0,2\}, \{1\}\}$ specifies that the kernels with ids 0 and 2 are mapped as a task component with both constituent kernels having `gpu' device preference while kernel with id $1$ is the other task component having `cpu' device preference. The number of command queues to be setup for the platform is specified in the list $cq$ where each element $r_i:n$ denotes that device $d_i \in \mathcal{P}$ has $r_i=n$ command queues. Assuming a target platform for 4 devices, we can observe from Fig. \ref{fig:json} that $cq=\{r_1:4,r_2:2,r_3:2,r_4:4\}$. The framework uses this information to automatically set up \textbf{$\mathcal{Q}$} data-structure for each task component. The dependency information of the DAG is specified as a set of edges of the form $k_i,b_r \rightarrow k_j,b_s$, where $k_i$,$k_j$ represent kernel ids that are dependent, $b_r$ is an output buffer of $k_i$ and $b_s$ is an input buffer of $k_j$ i.e. $(k_i,b_r) \in E_O$, $(b_s,k_j) \in E_I$ and $b_r,b_s \in E$. The ids for the buffers $b_r$ and $b_s$ are represented by their corresponding argument positions in the function call for the kernels. For example, consider the entry  $0,2 \rightarrow 2, 0$, in the {\tt dag.json} file of Fig. \ref{fig:json}. This implies that the output buffer specified in argument 2 of kernel $0$ will be used as input buffer specified in argument 0 of kernel $2$. Note in the bottom-left box of Fig. \ref{fig:json} the tag {\tt "outputBuffers"} for the {\tt matmul} kernel (which is kernel `0'). The argument position {\tt pos} for this buffer is indicated as 2 in the specification.

%As guidance parameters, the user can specify the device preference for each kernel using the {\tt dev} field. 
%The specification file of {\em PySchedCL} \st{consists of a collection of key-value pairs depicting} contains  necessary attribute information for an OpenCL kernel which includes information regarding input/output buffers, variables passed as arguments to the kernel call, the dimension of the kernel etc. It uses  Javascript Object Notation (JSON) file format. 
\vspace{-2mm}
	\vspace{-3mm}
 %The user has the option of either hard-coding buffer and work item sizes or writing expressions containing symbolic variables that depict the relationship between work items and the dataspace to be processed. This ensures that we have one specification file for a kernel and the final values of these symbolic variables can be provided as command line parameters at runtime. 
\begin{comment}
\begin{lstlisting}[caption={OpenCL Kernel for Matrix Multiplication},captionpos=b,frame=single,basicstyle=\tiny,language=C]
__kernel void gemm(__global float *A, __global float *B, __global float *C, 
int M, int N, int K) {
  int ty = get_global_id(1);
  int tx = get_global_id(0);
  if ((tx < N) && (ty < M)){
    C[ty * N + tx] = 0;
    for(int k=0; k < K; k++)
        C[ty * N + tx] += A[ty * K + k] * B[k * N +tx];
 }
}
\end{lstlisting}
\end{comment}
%We examine the specification information in detail for the matrix multiplication kernel shown in Fig. \ref{fig:json}. The corresponding kernel code is depicted in Listing 1. This is a 2-D kernel which takes as input two matrices $A$, $B$ of dimensions $M \times K$, $K \times N$ respectively and produces an output matrix $C$ of dimension $M \times N$. A total of $M*N$ work items is launched where the job of each work item is computing the dot product of one row of $A$ and one column of $B$ to produce one element of $C$. The required attribute information comprising information specific to the kernel implementation, buffers and variable arguments for the matrix multiplication kernel is depicted in Fig. \ref{fig:json}.	

Kernel information includes i) the name of the function ({\tt matmul}), ii) the filepath of the required source file ({\tt gemm.cl}), iii) the kernel dimensionality in {\tt workDimension} and iv) the total number of work items ({\tt globalWorkSize}) to be launched for this kernel. The variable {\tt globalWorkSize} is a three element list where each element refers to the number of work items along a particular dimension. Buffer information for each kernel constitutes information for three buffer lists - i) {\tt inputBuffers} reserved for input buffers, ii) {\tt outputBuffers} reserved for output buffers and iii) {\tt ioBuffers} reserved for buffers which are treated as both input and output  by the kernel. Each buffer is characterized by the tuple $\langle type, size ,pos \rangle$ where $type$ denotes the data type for each element in the buffer, $size$ denotes the total number of elements in the buffer, and $pos$ denotes the index position of the buffer argument in the actual function call of the kernel. %For example, the output buffer passed as argument in the third position of the function call in Listing 1 has $pos=2$. 
Variable argument information for every kernel is denoted by the tuple $\langle type,  pos, value\rangle$ with $type$, $pos$ meaning same as earlier and $value$ denoting the argument variable.  %The value of the $size$ parameter for buffers can be specified using compile time constants or symbolic expressions defined over the $value$ parameters of the kernel argument. 
\par\noindent The required attribute information comprising information specific to the kernel implementation, buffers and variable arguments for the {\tt matmul} kernel is depicted in Fig. \ref{fig:json}.  The {\tt matmul} kernel takes as input two matrices $A$, $B$ of dimensions $M \times K$, $K \times N$ respectively and produces an output matrix $C$ of dimension $M \times N$. For this, a total of $M*N$ work items is launched. In Fig. \ref{fig:json}, the three variable arguments are $M,N,K$ for {\tt matmul} and {\tt globalWorkSize = [M,N,1]}. In Fig. \ref{fig:json}, the value of $size$ for the output buffer is set as the symbolic expression $M*N$. Guidance parameters specified as symbolic expressions aide in depicting the relationship between number of work items launched and the dataspace to be processed.  The values of the symbolic variables $M, N, K$ can be configured by the user as command line parameters before dispatching the kernel. In general, guidance parameters can be  specified both as compile time constants or using expressions containing symbolic variables as exemplified. 
\par\noindent We have implemented an LLVM \cite{llvm} based compiler pass with the help of the library reported in \cite{coarsen} that 1) parses the abstract syntax tree of each OpenCL kernel and automatically generates necessary attribute information for each kernel, 2) infers  the dimensionality, types and positions of variables and buffers used for each kernel, 3) classifies buffers as input/output buffers by understanding whether it is treated as \textit{l-values} or \textit{r-values} in the body of the function. This pass thus automatically generates most fields of the JSON file by analysing the individual kernels. After this, the user is only required to specify guidance parameters which include -i) the size of the buffers ii) the number of work items iii) the values of the variable arguments.

%where $type$ denotes the type of the variable, $value$ represents the value contained in the variable and $pos$ represents the index position of the variable argument in the actual function call of the kernel. \ag{Need to keep information about variable arguments. The values of M,N,K would be required internally in the source code to calculate index expressions.}

%\par \sd{As opposed to frameworks like SOCL, StarPU and MultiCL which requires designing host-side implementations, our framework relies only on the DAG specification provided as a simple JSON file containing dependency information and kernel attribute information. In the process, considerable implementation overhead is reduced for the user. -- REPITITION ???} \ag{Yes this can be removed}
\par \noindent \textbf {B. Scheduling Backend: } As depicted in Fig. \ref{fig:pyschedcl}, the scheduling backend processes the specification file of an application DAG and takes care of setting up OpenCL command queues for mapping kernels across devices of a heterogeneous CPU/GPU platform. We explain the working principle of the backend with the help of the procedure $schedule$ highlighted in Algorithm \ref{algo:dispatch}. 
\vspace{-3mm}
\begin{algorithm}[ht]
		\caption{Scheduling in PySchedCL\label{algo:dispatch}}
		{\scriptsize 
		\textbf{Input: } $G$ - an OpenCL DAG, $\mathcal{P}$ - target devices
        \begin{algorithmic}[1]
		\Procedure{schedule}{$G$,$\mathcal{P}$}
			\State $\mathcal{F} \leftarrow ready\_task\_components(G)$ , $\mathcal{A} \leftarrow \mathcal{P}$
			\While{all kernels of $G$ not finished}
			\While{$\mathcal{A}$ contains a device and $\mathcal{F}$ is not empty}
			\State $T,d \leftarrow select(\mathcal{F},\mathcal{A})$, \textbf{$\mathcal{Q}$} $\leftarrow setup\_cq(T,d)$, $dispatch(T$,\textbf{$\mathcal{Q}$}$)$
			\EndWhile
			\State $sleep\_till\_cb\_update()$
			\EndWhile
			\EndProcedure
		\Function{setup\_cq}{$T$,\textbf{$\mathcal{Q}$}} 
			\State \textbf{$\mathcal{Q}$}= $\langle Q,E_Q \rangle \leftarrow init()$ $unprocessed \leftarrow FRONT(T)$
			\While{all kernels of $T$ not processed}	\State $k \leftarrow unprocessed$, $q \leftarrow sel_{rr}(Q)$, $enq(k,q)$
			\State $set\_dependencies(k,E_Q)$,  $update(unprocessed)$
			\IfThen{$k \in END(T)$}{$set\_callbacks(k,$\textbf{$\mathcal{Q}$}$,cb)$}
			\EndWhile
			\EndFunction
			\Function{cb}{$ $} 
			\State $ev,T,d, \mathcal{F}, \mathcal{A} \leftarrow get\_user\_args()$
			\State $update\_status(ev,T)$, $\mathcal{T}^\prime = get\_ready\_succ(T)$
			\State $lock()$; $update\_task\_queue(\mathcal{T}^\prime,\mathcal{F}$); $unlock()$
			\IfThen{$T$ is finished}
		{$lock()$;$return\_device(d,\mathcal{A}$); $unlock()$}
		\EndFunction
		\end{algorithmic}
		}
	\end{algorithm}
	\vspace{-3mm}
	 %The input to $schedule$ is the application graph $G$ and the set of devices in the target platform $\mathcal{P}$ in the JSON file.
\par \noindent \textbf{Initialization: }The procedure executes on the host device, and first parses the input specification (for the application graph $G$ along with the set of devices in the target platform $\mathcal{P}$) and populates the centralized task queue $\mathcal{F}$ with task components that are ready for dispatch using $ready\_task\_components()$ (line 2). Here, a task component $T$ is added to $\mathcal{F}$ if for every kernel $k_i \in FRONT(T)$, there exists no predecessor. % i) there exists no predecessor or ii) all predecessors of $k_i$ have finished execution.
The task queue $\mathcal{F}$ is implemented as a priority queue, where the user can specify custom ranking measures for enforcing an ordering among task components to be selected from $\mathcal{F}$.
%We note that task components can be individual kernels or a collection of kernels depending on the value of the $tc$ flag in the specification file. For the former case, scheduling decisions are coarse-grained in nature i.e. all the operations associated with a kernel $k_i$ are finished before proceeding to execute a successor kernel $k_j$. 
The set $\mathcal{A}$ represents the set of available devices and is initialized to all the devices contained in $\mathcal{P}$ (line 2). 
\par\noindent \textbf{Primary Scheduling Loop: }The procedure $schedule$ runs the routines inside the while loop (line 3-6) and continues until all kernels of the DAG have not finished execution. In any scheduling iteration where the frontier $\mathcal{F}$ and the set $\mathcal{A}$ are non-empty (line 4), the $select$ routine inspects task components in $\mathcal{F}$ and returns $T$ and $d$ (line 5) if an available device $d \in \mathcal{A}$ is found that matches the device preferences of all the constituent kernels of some task component $T \in \mathcal{F}$. We note that the $select$ routine is a blocking call i.e. if a task component $T$ or a  matching device $d$ is not immediately available, the routine blocks further execution of the $schedule$ routine until a suitable match  is found. In the meantime, kernels already dispatched continue executing on their respective devices. Once $T$ and a matching device $d \in \mathcal{A}$ is obtained using $select$(line 5), the framework spawns a separate child thread responsible for running $setup\_cq()$ and $dispatch()$ functions for mapping $T$ to $d$ (line 5). This ensures that on the host device i) the master thread running $schedule$ continues to search for existing task components that are free to execute on matching devices that are available using $select$ (lines 3-5) and ii) the subsequent operations  for setting up  command queue data structures (once a match is found) and dispatching each task component (line 5) are performed in parallel using separate child threads. In the event, if one of $\mathcal{F}$ or $\mathcal{A}$ is empty, the $schedule$ procedure remains idle using $sleep\_till\_cb\_update()$ (line 6). The data structures $\mathcal{F}$ and $\mathcal{A}$ are updated by callback functions with new task components and devices, once they are available again. This happens following the procedure outlined in $cb$ (lines 13-17). We note such callback functions when initiated operate in a parallel thread w.r.t. $schedule$, executing on the host device. Once one of these data structures are updated by respective callbacks, the $schedule$ routine resumes the scheduling loop (lines 3-5) if any kernel of $G$ remained unfinished.  
\par
\begin{figure}[ht]
		\centering
		\includegraphics[scale=0.4]{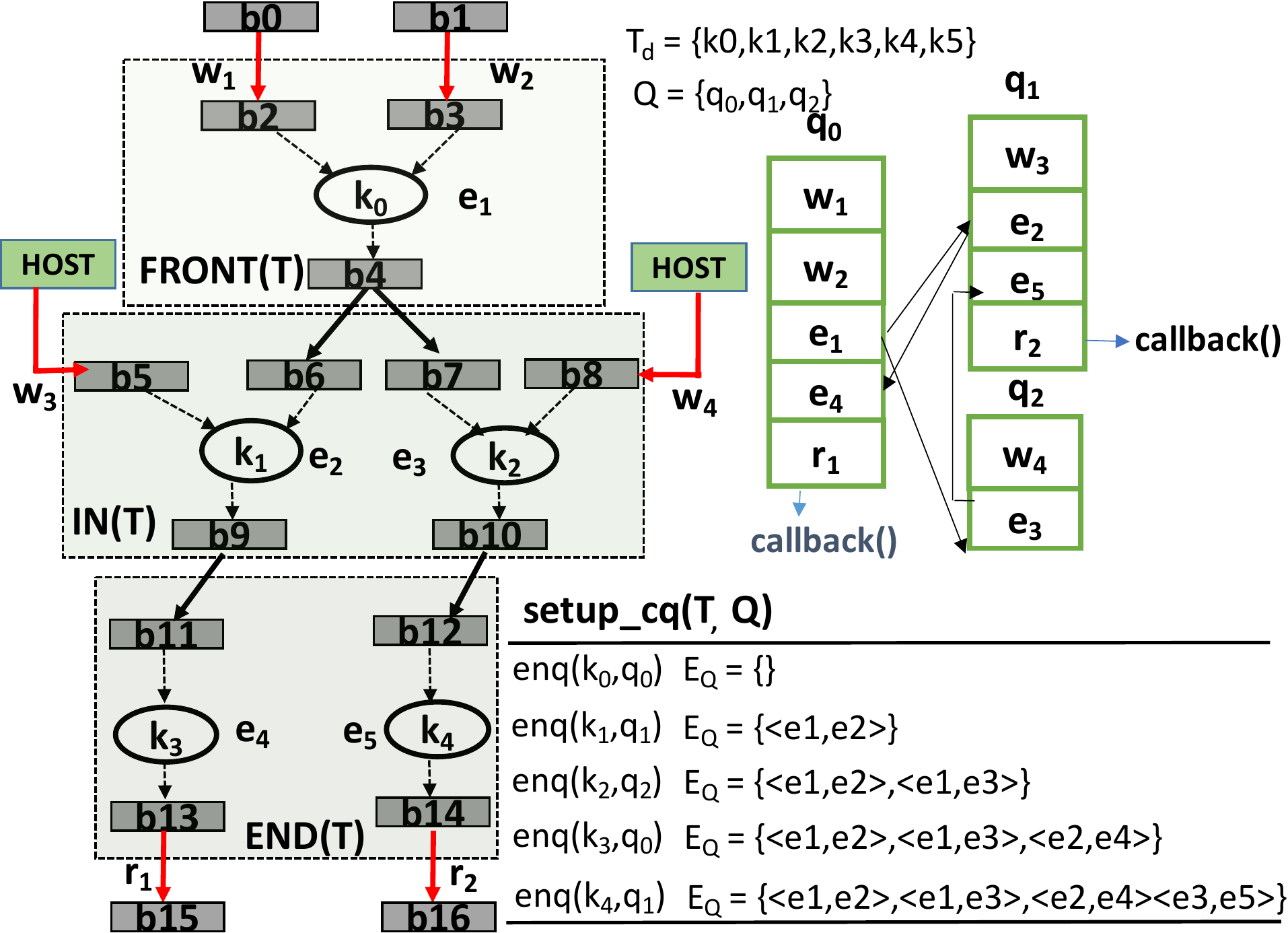}
		\vspace{-2mm}
		\caption{\small Command Queue Setup\label{fig:dispatch}}
	\end{figure}

 \noindent \textbf{Command Queue Setup: } In each child thread, the $setup\_cq$ procedure is used to set up the command queue structure \textbf{$\mathcal{Q}$} (line 5) for a given $T$ and $d$. The data structure \textbf{$\mathcal{Q}$} $= \langle Q,E_Q \rangle $ is first initialized using the $init$ routine (line 9) such that $E_Q = \{ \}$ and $Q=\{q_0,q_1,\cdots,q_{D}\}$ where $D$ represents the number of command queues to be setup as specified by $r_i \in cq$ for that device $d$ in the JSON file. We shall explain how $setup\_cq$ finishes setting up \textbf{$\mathcal{Q}$} with the help of an illustrative example depicted in Fig. \ref{fig:dispatch} where we map the task component $T = \{k_0,k_1,k_2,k_3,k_4 \}$ to a GPU device using a total of 3 command queues i.e. $Q = \{q_0,q_1,q_2 \}$.  
	\par \noindent After $init()$, the procedure next initializes the set $unprocessed$ with kernels belonging to $FRONT(T)$ (line 8) and keeps on updating \textbf{$\mathcal{Q}$} until all kernels of $T$ have been processed (lines 9-12). A kernel $k$ is said to be processed once all read, write and ndrange operations pertaining to it have been enqueued to a queue, otherwise it is unprocessed. Initially, $unprocessed = \{k_0\}$ for the task component $T$ in Fig. \ref{fig:json}. Next, in each iteration of the while loop (lines 10-12) a kernel $k$ is first selected from  $unprocessed$ (line 10). A queue $q$ is selected in a round robin fashion from $Q$ of \textbf{$\mathcal{Q}$} using $sel_{rr}(Q)$ (line 10).  Following the rules outlined in Section \ref{sec:prob}, the $enq(k,q)$ function enqueues relevant read, write and ndrange commands of kernel $k$ to $q$ (line 10). 
	In Fig. \ref{fig:dispatch}, we observe that $k_0$ and $q_0$ are first selected and $enq(k_0,q_0)$ pushes write commands $w_1$ and $w_2$ to $q_0$ followed by the ndrange command $e_1$. The write commands correspond to the two inter edges $(b0,b2)$ and $(b1,b3)$. The $setup\_cq$ function next sets up dependencies between relevant operations i.e. synthesizes $E_Q$ of \textbf{$\mathcal{Q}$} using $set\_dependencies()$ (line 11). For kernel $k_0$, we have no dependencies to set. Once this is done, the list $unprocessed$ is updated with the successors of $k$ that have not been processed using $update$ (line 11). 
	\par The sequence of $enq$ calls and step by step construction of the set $E_Q$ by $setup\_cq$ are highlighted in Fig. \ref{fig:dispatch}. For kernel $k_1$ belonging to $IN(T)$, $enq(k_1,q_1)$ pushes the isolated write operation $w_3$ and the ndrange operation $e_2$ to $q_1$. The $set\_dependencies()$ function populates $E_Q$ with the dependency $\langle e_1,e_2 \rangle$. One may observe that despite there being a dependency between $k_0$ and $k_1$, the round-robin selection of queues ensures that write commands $w_1$ and $w_2$ enqueued to $q_0$ can be interleaved with the write command $w_3$ enqueued to $q_1$. In a similar fashion, as depicted in Fig. \ref{fig:dispatch}, the $enq$ function  pushes operations for kernels $k_2$, $k_3$ and $k_4$ to $q_2$, $q_0$ and $q_1$ respectively while the $set\_dependencies()$ function sets up $E_Q$. One can also observe that since ndrange operations $e_2$ and $e_3$ belong to different command queues with no dependencies, they can also execute in parallel. We note that while setting up $\mathcal{Q}$, the framework uses low-level OpenCL API calls with the {\tt clEnqueue} prefix for i) enqueuing commands in each $q\in Q$ ii)  associating event objects with each such command and  iii) enforcing dependencies in $E_Q$ using these event  objects as discussed in Section \ref{sec:OpenCL}. 

\par \noindent \textbf{Callback Assignment: } In  addition to constructing  \textbf{$\mathcal{Q}$}, $setup\_cq$ uses the $set\_callbacks$ function (line 12) to investigate already enqueued operations pertaining to each kernel $k\in END(T)$ in $\mathcal{Q}$ and register multiple instances of the callback procedure $cb$ (lines 13-17). This is done by registering an instance of $cb$ using {\tt clSetEventCallback()} (refer Section \ref{sec:OpenCL}) for every event $ev$ associated with certain commands for all kernels $k \in END(T)$ in $\mathcal{Q}$, depending on the device $d$ where $T$ gets  mapped to. 
	\par\noindent 1) If $d$ is a GPU device, callback is registered for events associated with every dependent \textit{read} command pertaining to an inter edge $b_i,b_j \in E$ such that $k, b_i\in E_O$. In Fig. \ref{fig:dispatch}, callbacks are registered for events associated with the read commands  $r_1$ and $r_2$ pertaining to kernels $k_3$ and $k_4$ belonging to $END(T)$.
	\par\noindent 2) If $d$ is a CPU device sharing the same memory space as that of the host, callback is registered for the event pertaining to the \textit{ndrange} operation of $k$ if $k,b_i\in E_O$ and there exists an inter edge $b_i,b_j \in E$. If $T$ had been mapped to a CPU device in Fig. \ref{fig:dispatch}, the callbacks would have been registered with events associated with the ndrange commands $e_4$ and $e_5$.
	\par While registering each callback instance, the routine $set\_callbacks()$ also ensures to specify the associated event $ev$, task component $T$, device $d$ and global task queue $\mathcal{F}$ and device set $\mathcal{A}$  in $user\_arg$ argument of {\tt clSetEventCallback()} so that they are accessible once the callback instance starts execution.
	\begin{comment}
	\par \noindent \textbf{Callback Assignment: } In addition to constructing  $\textbf{\mathcal{Q}}$, $setup\_cq$ uses the $set\_callbacks$ function (line 12) to register the callback function $cb$ (lines 13-17) for events pertaining to every kernel $k \in END(T)$. This is done using the {\tt clSetEventCallback()} as discussed in Section \ref{sec:OpenCL} for events associated with the read commands  $r_1$ and $r_2$ pertaining to kernels $k_3$ and $k_4$ belonging to $END(T)$ in Fig. \ref{fig:dispatch}. 
	
	In general,  callback functions are setup for dependent read commands for every inter edge $(b_i,b_j)$ such that $(k,b_i)\in E_O$ and $k \in END(T)$. 
	
	Since $k_3$ and $k_4$ belong to $T$ mapped to a GPU device and the successors of these kernels belong to a different task component mapped to a different device, the buffers $b_{13}$ and $b_{14}$ need to get copied back to the host. If $T$ had been mapped to a CPU device sharing the same memory space as that of the host, the read operations would have been redundant. In this scenario, the callbacks would be associated directly with ndrange commands $e_4$ and $e_5$. The routine $set\_callbacks()$ also ensures using {\tt clSetEventCallback()} such that each callback function $cb$ has access to the associated event $ev$, task component $T$, device $d$, task queue $\mathcal{F}$ and device set $\mathcal{A}$. 
	\end{comment}
	\par \noindent \textbf{Thread safe Callback Procedure: } Each instance of the callback function registered using $set\_callbacks()$ is spawned at runtime when the associated event completes and follows the functionality outlined in procedure $cb$ (lines 13-17). The associated data $ev$, $T$, $d$, $\mathcal{F}$ and  $\mathcal{A}$ as discussed above are first obtained using $get\_user\_args()$ (line 14). Next, the routine $update\_status$ (line 15) is used to update based on $ev$ which kernel $k_i$ has completely finished execution in $END(T)$. We say that a kernel $k_i$ has finished execution if i) $k_i$ was mapped to a CPU and $ev$ pertained to an \textit{ndrange} command or ii) $k_i$ was mapped to a GPU and every event other than $ev$ pertaining to dependent \textit{read} commands have also completed. This indicates that the output buffers produced by $k_i$ are available in the host memory space. Depending on which kernel $k_i$ has finished, task components $T^\prime \notin \mathcal{F}$ containing kernels $k_j \in FRONT(T^\prime)$ which are successors of $k_i$, are next investigated if they are ready for dispatch in the function $get\_ready\_succ()$ (line 15). If it is observed that all predecessors of every kernel in $FRONT(T^\prime)$,  have finished execution, then $T^\prime$ is ready for dispatch. All such ready task components are populated in the set $\mathcal{T}^\prime$ and are added to $\mathcal{F}$ using  $update\_task\_queue()$ (line 16). Finally, if all kernels of $END(T)$ have finished execution i.e. all kernels in $T$ have completed, the device $d$ is returned back to $\mathcal{A}$ using the $return\_device$ function (line 17). One may further note that the routines $return\_device()$ and $update\_task\_queue()$ are rendered thread safe using the $lock()$ and $unlock()$ functions. As discussed earlier, callback functions are initiated in separate threads and thus execute in parallel with the host thread running $schedule$ while potentially modifying $\mathcal{F}$ and $\mathcal{A}$. Furthermore, as described above  multiple callbacks can be registered for the same task component and can potentially execute simultaneously. It is therefore imperative that atomic updates are applied to the shared data structures $\mathcal{F}$ and $\mathcal{A}$ by the callback functions to ensure correctness.  
	%We note for coarse-grained scheduling decisions,  all relevant operations will be enqueued to one single command queue in $\textbf{\mathcal{Q}}$.
	\par \noindent \textbf{Final Dispatch: } Once the callback functions are set, the {\tt dispatch} function (line 8) is called which executes the {\tt clFlush()} function once for each command queue in $Q$ to ensure that the commands are submitted to the device. Once this call is made, the associated command queues are locked i.e. they cannot be used by other task components that are ready for dispatch. 
%\par The algorithm $schedule$ highlights a generic procedure for scheduling algorithms in our framework. By specifying the task component partitioning $\mathcal{T}$, and overriding the implementation of the $select$ routine for choosing $T \in \mathcal{T}$, and $d\in\mathcal{A}$, the user can experiment with different scheduling policies. We implement one fine-grained scheduling algorithm and two coarse-grained scheduling algorithms and present experimental results for the same, continuing with our  motivational example from Section \ref{sec:OpenCL}.
\par The algorithm $schedule$ highlights a generic scheduling framework %where the $select$ routine chooses task component $T \in \mathcal{T}$, and device  $d\in\mathcal{A}$ in every scheduling iteration using the strategy as described. The framework 
which  allows for specifying the task component partitioning $\mathcal{T}$ and overriding the implementation of the $select$ routine for choosing $T \in \mathcal{T}$  and $d\in\mathcal{A}$ with different scheduling policies.
We note that the design principles of {\em PySchedCL} extend beyond OpenCL and holds in general for any heterogeneous CPU/GPU platform that supports SIMD style programming and a runtime system that supports some abstraction of worker queues  for enqueuing operations on compute devices. One can thus incorporate the methodologies highlighted in this paper by implementing $schedule$ in CUDA (using {\em streams}) or modern frameworks such as DPC++ (using {\em SYCL queues} \cite{dpc}). The focal point of our work lies in investigating fine-grained scheduling techniques as opposed to traditional coarse-grained policies and presenting a generic design workflow for achieving the same.  %\sd{We implement  one fine-grained and two coarse-grained scheduling  algorithms overriding the baseline $select()$ and present experimental results for the same, continuing with our  motivational example from Section \ref{sec:OpenCL}. - DELETE IF REQ}
\section{Experimental Results}
\label{sec:expt}%10 pages 
	
	%We consider a target platform comprising an NVIDIA GTX-970 GPU card and a Quadcore Intel i5-4690K CPU which was also used for our motivational example. 
	The Transformer Neural Network \cite{DBLP:journals/corr/VaswaniSPUJGKP17} has proven to be a viable alternative to Recurrent Neural Networks \cite{hochreiter1997long}, in Natural Language Processing (NLP) tasks such as Named Entity Recognition and Neural Machine Translation.  
	\par
\begin{figure}[ht]
		\centering
		\includegraphics[scale=0.3]{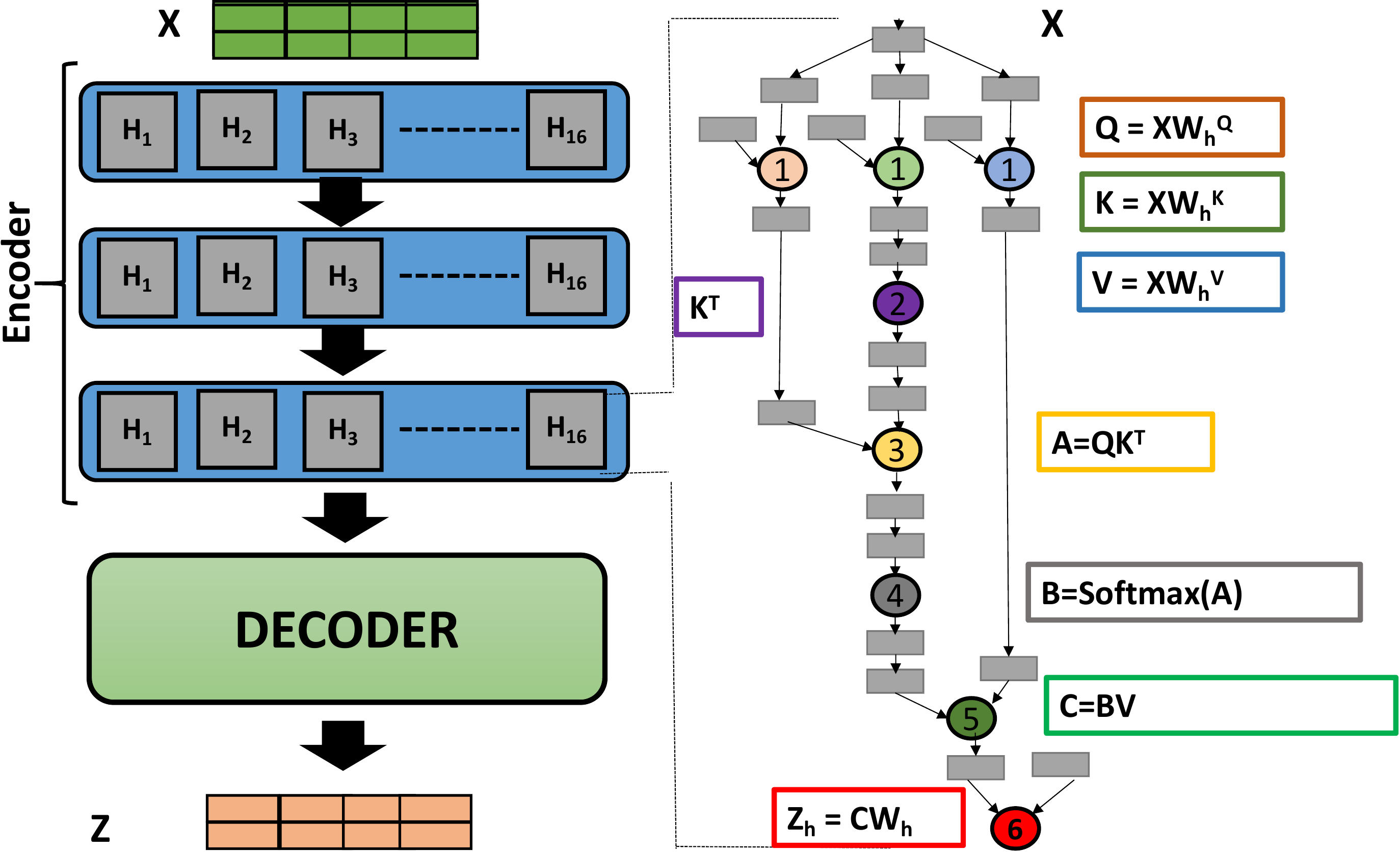}
		\vspace{-2mm}
		\caption{\small Transformer Architecture\label{fig:transformer}}
	\end{figure}

 The transformer architecture is based on the standard encoder-decoder architecture used in sequential learning tasks, and is depicted in Fig. \ref{fig:transformer}.	The input to the transformer is a sentence matrix $X =[w_{1}^\intercal,w_{2}^\intercal,w_{3}^\intercal....w_{n}^\intercal]$ where $w_i \in \mathbb{R}^d$ represents an embedding vector for each word in the sentence. The matrix $X$ undergoes transformations through each layer in the encoder and decoder before yielding the target vector $Y$. An {\em attention mechanism} is used to assign scores indicating the importance of each word in the sentence. This is achieved by a series of matrix transformations through a mechanism called multi-headed attention. Each layer in the transformer comprises multiple heads operating in parallel where each head $h$ represents a series of linear algebra operations on the sentence matrix $X$ for generating a contextual embedding matrix $Z_h$ comprising contextual embedding vectors for each of the $n$ words in the sentence. Each head $h$ is characterized by four parameter weight matrices $W_h^Q,W_h^K, W_h^V $and $W_h$. The computation involved in each head $h$ is represented by the DAG on the right hand side of Fig. \ref{fig:transformer} (the same DAG used in the motivation example in Section \ref{sec:OpenCL}). It can be observed that the sentence vector $X$ typically undergoes 3 parallel GEMM transformations with the weight matrices $W_h^Q,W_h^K, W_h^V $ to generate Query $Q$, Key $K$ and Value $V$ matrices respectively. Using $Q$ and $K$, as depicted in Fig. \ref{fig:transformer}, the matrices $A=QK^\intercal$ followed by $B=Softmax(A)$ are computed where $Softmax$ represents a normalized exponential function \cite{bishop}. The contextual embedding matrix $C = [h_{1}^\intercal,h_{2}^\intercal,h_{3}^\intercal....h_{n}^\intercal]$ is computed as $C = BV$. Finally, the output $Z_h$ is obtained by the GEMM operation $CW_h$. The outputs of each of these heads are concatenated to produce the final contextual embeddings for the sentence. The output of each layer is passed as input to the following layer similar to any neural network  pipeline.

	%\par In the recent past, Transformer \cite{DBLP:journals/corr/VaswaniSPUJGKP17} neural networks have proven to yield signficantly better results than the state of the art Recurrent Neural Network (RNNs) architectures such as LSTMs \cite{hochreiter1997long}, where given a sentence $ S = \{w_1,w_2,w_3....,w_n\} $, and $\forall i \in [1,n]$  \cite{NIPS2013_5021} \cite{Pennington14glove:global} of the sentence, the RNNS produced  context aware embeddings $h_i$ in a recursive manner as follows.
	%$$
	%h_i = f(h_{i-1},w_i;\theta)
	%$$
	%$$
	%h_1 = \vec{0}
	%$$
	%The recursive formulation of $f$ enforces that $h_{i}$ can only be computed once the set of embeddings $\{h_0,h_1,....h_{i-1}\}$ have been computed. 
\par \noindent In contrast to RNNs, the transformer architecture offers ample scope for parallelization. A single transformer layer typically comprises, 8 or 16 heads and thus a maximum of $16*3=48$ matrix computations can execute in parallel given sufficient hardware resources. Since the operations in each of the layers are similar in nature, we validate our experimental findings by designing a single layer of the transformer network as a DAG using a {\em PySchedCL} based  specification. As component kernels, we use kernels that are readily available from the Polybench \cite{polybench}, NVIDIA OpenCL \cite{nvidia} benchmark suites. 
	\par We conduct a series of experiments classified into three broad categories and for each experiment we define $\beta$ as the size of the transformer such that the matrices defined earlier $Q$,$K$,$V$,$X$ are all of dimensions $\beta \times \beta$. We denote the number of heads for the transformer as $H$. All experiments are executed on the single CPU single GPU platform described in Section \ref{sec:OpenCL}.
\par \noindent \textbf{Experiment 1: Static Scheduling Scheme for Transformer}: We implement a static scheduling scheme called {\em clustering} which is optimized for taking fine-grained decisions  and profile the execution time for a total of 16 transformer DAGs.  The 16 DAGs for the experiment set are generated by varying the number of heads $H \in [1,16]$ and fixing $\beta$ for each DAG as 256.  
\par For our {\em clustering} scheme, we set the task component mappings for each of the 16 DAGs beforehand by configuring i) the kernel preferences using the $dev$ field for each kernel and ii) the task component partitioning $\mathcal{T}$ using the list $tc$ in the specification file. Given the structure of the DAG, it makes sense to cluster all kernels belonging to one transformer head into a task component and map it to a particular device. Since, the transformer heads are independent, such task component mappings would result in  there being no inter edge buffers. As a result there will be no read callbacks. Thus for any transformer with $H$ heads, possible mapping configurations would be to 1) map all heads to a GPU device, 2) map 1 head to the CPU and $H-1$ heads to the GPU device, ... and finally $H+1$) mapping all $H$ heads to the GPU device.  Since each head is identical, clustering all kernels of a head into a task component would result in a total of $H+1$ mapping configurations for a DAG with $H$ heads.
\par In the {\em clustering} scheme, each task component $T$ for the DAG is annotated with the maximum bottom level rank \cite{heft} of the kernels in $FRONT(T)$. The bottom level rank for any kernel in a DAG represents the maximum time left to finish all kernels in the path starting from $k$ to the last kernel in the DAG. The priority queue $\mathcal{F}$ is ordered with respect to this bottom level rank measure. Since the order of kernel dispatching and device mapping decisions are decided beforehand, the {\em clustering} scheme is inherently static. 
\par We execute the {\em clustering} scheme for each of the transformer DAGs by varying the number of command queues $q_{cpu}\in [0,5] $ and $q_{gpu} \in [0,5]$ for our target platform and by varying the number of task components mapped to the CPU, $h_{cpu} \in [0,H]$.  The remaining $H-h_{cpu}$ task components are mapped to the GPU device. Given this, let us denote an architecture mapping configuration for the {\em clustering} scheme as $mc=\langle q_{gpu},q_{cpu},h_{cpu}\rangle$. We have observed from our experimental results that increasing beyond 5 command queues for the CPU and GPU device does not improve execution time. This may be attributed to the overhead of managing multiple command queues by the OpenCL runtime system. The {\em clustering} scheme can also emulate static coarse-grained scheduling decisions if the entire DAG is mapped to the GPU device using a single command queue i.e. $mc=(1,0,0)$. We consider {\em clustering} with this architecture-mapping configuration to be the default coarse-grained scheduling scheme against which we shall compare how fine-grained decisions fare for the same task component partitioning $\mathcal{T}$.
\par
\begin{wrapfigure}[10]{L}{0.25\textwidth}
	%\begin{figure}[ht]
		\centering
		\includegraphics[trim = 0 0 22 10, clip, width=0.25\textwidth]{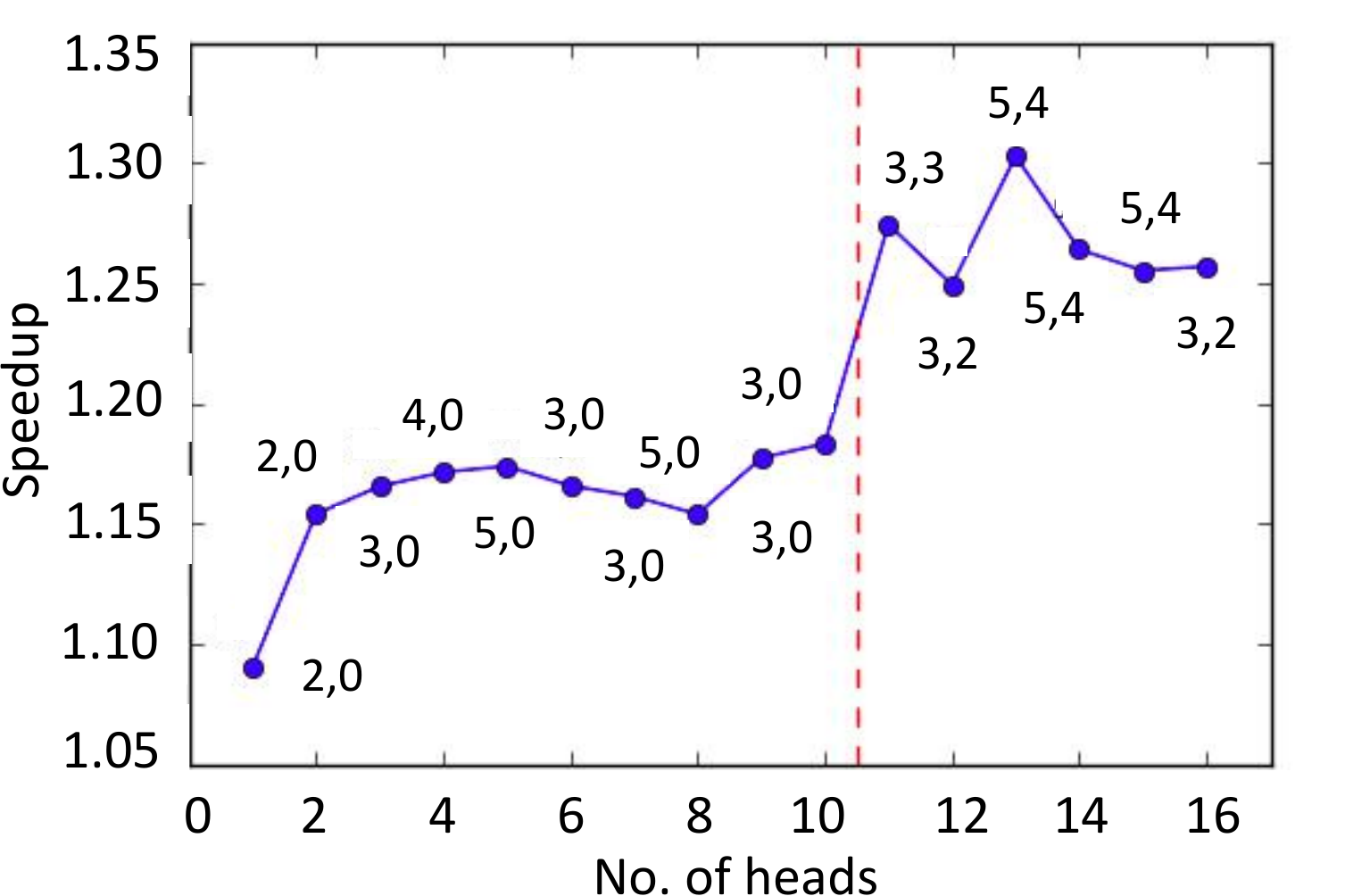}
		\vspace{-6mm}
		\caption{\small Speedups for Expt. 1 \label{fig:Expt1}}
	%\end{figure}
	\end{wrapfigure}

 For each DAG distinguished by the number of heads $H$, the best mapping configuration for {\em clustering} is the one which gives the best speedup with respect to time taken by the DAG to execute in its default configuration. We profile a total of $(H+1)*q_{cpu}*q_{gpu}$ such mapping configurations for each transformer DAG with $H$ heads and highlight our observations in  Fig. \ref{fig:Expt1}. The x-axis denotes the total number of heads for the transformer. The y-axis represents the speedups obtained for the best configuration for each DAG over the default configuration. Each point is labeled by the $q_{gpu},q_{cpu}$ tuple corresponding to the best configuration. We further note that for DAGs with number of heads upto 10 (region to the left of the dotted line), $h_{cpu}$ is 0. For DAGs having  number of heads greater than 10 (region to the right of the dotted line), we have $h_{cpu}=1$. 
\par \noindent Thus, for DAGs with $H \in [1,10]$, we observe that the best configuration only differs from the default configuration with respect to the $q_{gpu}$ parameter. All the task components of the DAGs are scheduled to the GPU with the only difference being the number of command queues assigned to each component. The key observation in this region is that the transformer shows a clear speedup of about $15\% - 17\%$, if fine-grained scheduling is enabled leveraging multiple command queues. This highlights the effectiveness of automated fine-grained scheduling which our framework offers.
\par For $H \in [11,16]$, we observe that scheduling one of the task components of the DAG to the CPU device yields the maximum speedups.  We also observe a jump in the relative speedup values as compared to the DAGs with $H <=10$. This is because apart from taking fine-grained scheduling decisions for the GPU device, we are also undertaking certain fine-grained scheduling decisions for the CPU device as well. This results in better extraction of application level-parallelism since mapping a task component to the CPU results in lesser contention for the GPU device. However we observe with $H > 10$, it is meaningful to migrate only one head to the CPU device for further speedups.  This makes sense since i) the GPU has an order of magnitude number of processing elements greater than the CPU under consideration, ii) the kernels selected are optimized for GPUs rather than CPUs and iii) the CPU device is heavily engaged in setting up command queues and issuing directives to the devices in the heterogeneous platform.  Mapping more than 1 head for execution on the CPU actually takes more time to execute than that of mapping the remaining heads to the GPU device.
	%\par We observe two key points from this experiment. Firstly, the transformer DAG is mapped efficiently using fine-grained scheduling decisions with the help of setting up multiple command queues, thereby lending credence to our framework's central idea of exploiting concurrency. Secondly, we observe that only for DAGs with $H > 10$, it is meaningful to utilise the CPU device for further speedups.  
	\par The current experiment highlighted how the {\em clustering} scheme has been envisaged for the transformer DAG by considering specific DAG head mappings along with a choice of command queues and devices. Our next two categories of experiments consider comparing our static scheme with traditional implementations of coarse-grained heterogeneous scheduling policies like Eager and HEFT available in the StarPU framework \cite{augonnet2011starpu} which are inherently dynamic in nature i.e. kernel-device mappings are decided only at runtime.
	\par \noindent \textbf{Experiment 2: Clustering vs Eager Execution}:  We have implemented a simplistic {\em eager} execution based scheduling algorithm in our framework  inspired from StarPU. In this strategy, we specify  every kernel in the DAG as a separate task component and each device to use only one command queue. The $select$ routine is modified to i) choose task components based on the bottom level ranks discussed earlier and ii) select any device $d$ that is available at runtime irrespective of the individual device preferences of the kernel. The eager scheduling scheme supports only coarse-grained scheduling since it implements single command queue per device. Furthermore, since each task component is a kernel here, an explicit callback is required for every kernel to notify the host that it has finished execution.
 	\par
\begin{figure}[ht]
		\centering
		\includegraphics[trim = 32 0 0 0, clip, scale=0.34]{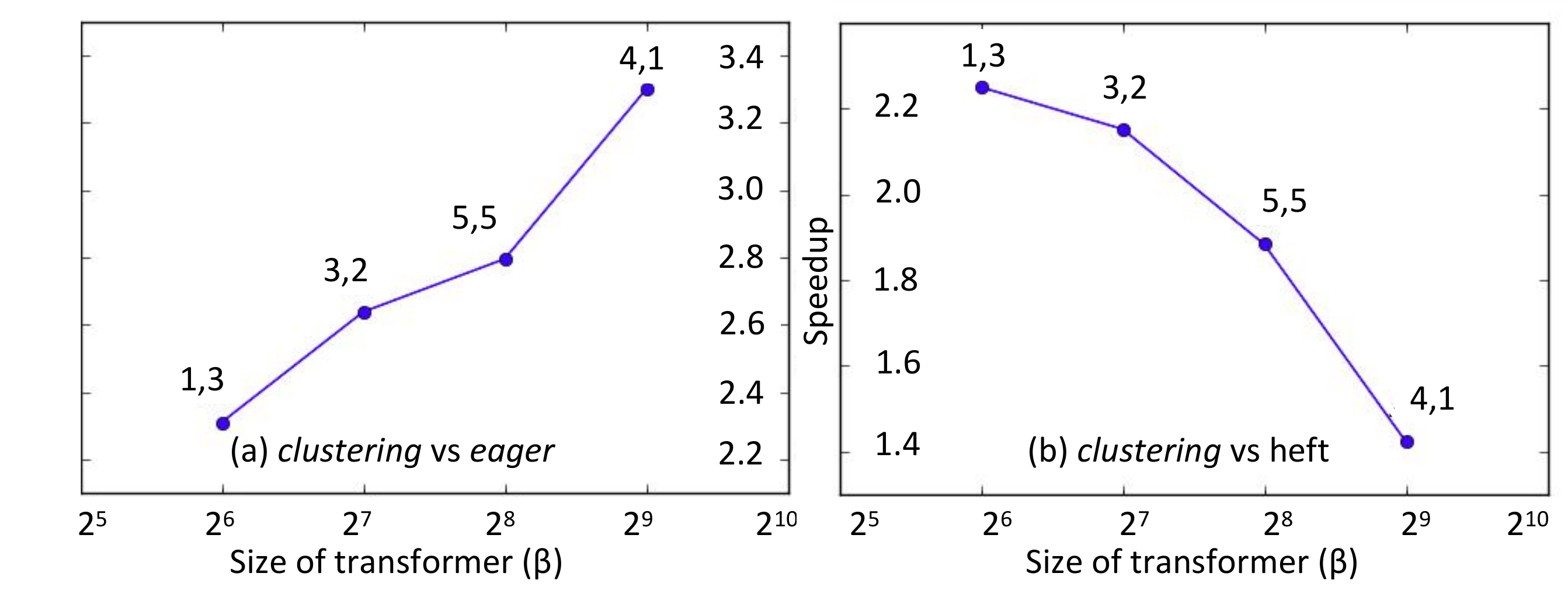}
		\vspace{-6mm}
		\caption{\small Speedup Results for Experiments 2 and 3 \label{fig:Expt2n3}}
	\end{figure}

 For a comparative evaluation of this dynamic scheme with {\em clustering}, we generate a set of input DAGs by keeping the number of heads $H$ fixed to $16$ and by varying the size parameter $\beta$ from $64$ to $512$ in powers of 2. We profile each such input DAG, using both {\em eager} scheduling and {\em clustering} schemes for all possible mapping configurations. We compute the speedups of execution times taken by {\em clustering} based scheduling for the best mapping configuration over that of {\em eager} and highlight them in Figure \ref{fig:Expt2n3} (a). The x-axis represents the size of the transformer head ($\beta$) for each DAG and the y-axis represents the speedup values. Each point in the plot is again labelled by the tuple $q_{gpu},q_{cpu}$ used by the best mapping configuration for the {\em clustering scheme}. The third element of the best configuration $h_{cpu}$ was found to be 1 for each $\beta$. It can be observed that {\em clustering} outperforms {\em eager} by a considerable margin. The comparison in Fig. \ref{fig:Expt2n3} brings out the advantage of static fine-grained scheduling as supported by our framework.
	 \par \noindent \textbf{Experiment 3: Clustering  vs HEFT:} Our final experiment considers the standard Heterogeneous Earliest Finishing Time First algorithm \cite{heft} {\em heft}.  Similar to {\em eager}, the scheduling heuristic {\em heft} assumes each task component to represent one kernel and sets up one command queue for each device. The  $select$ routine is modified to i) choose the kernel $k$ with the maximum bottom level rank and for ii) choose the device $d$ on which $k$ can execute with the earliest finishing time (EFT). Assuming execution times for kernels are available via prior profiling, EFT of $k$ executing on a device $d$ is computed as the sum of its execution time and the execution time of a kernel $k^\prime$ currently executing on $d$. %Note, our implementation for HEFT is not calibrated to take into account scheduling overheads such as enqueuing delays while selecting devices. 
	Considering the same set of input DAGs used in Experiment 2, we plot the speedups of the best configurations of the {\em clustering} scheme over the {\em heft} scheme in  Fig. \ref{fig:Expt2n3} (b). As expected, {\em heft} performs better than {\em eager} due to the added knowledge of earliest finishing times for each task. However, {\em heft} being implemented as a dynamic coarse-grained scheduling scheme is short sighted and fails to exploit concurrency aware scheduling decisions undertaken by {\em clustering}.
	\par
\begin{figure}[ht]
	\centering
	\includegraphics[scale=0.47]{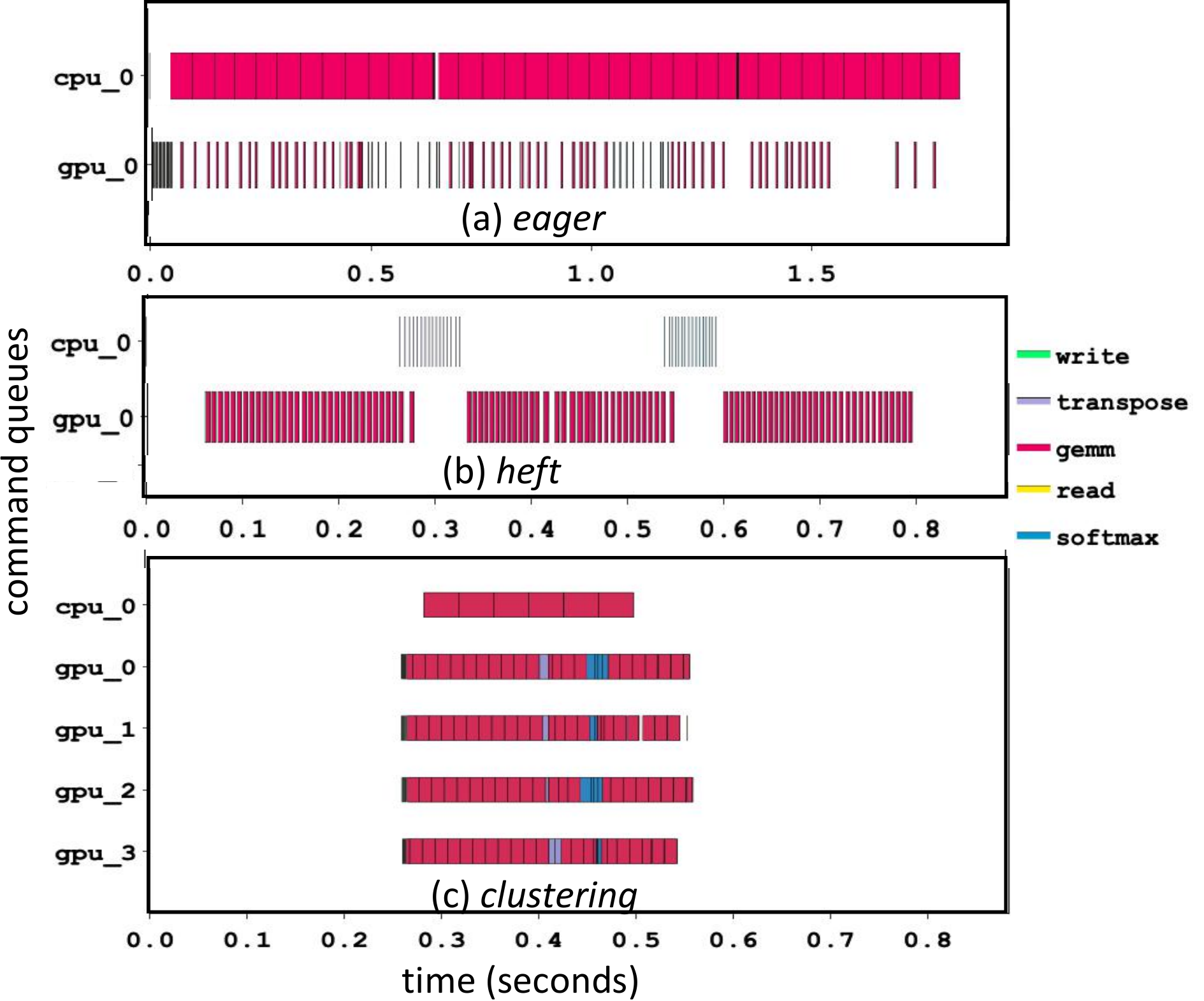}
	\vspace{-4mm}
	\caption{\small Gantt charts for different scheduling algorithms \label{fig:combined_gannt}}
\end{figure}

 \noindent \textbf{Comparative Evaluation:} One can also observe from the speedup values from Experiments 1, 2 and 3 that both {\em heft} and {\em eager} perform poorly when compared to both coarse-grained and fine-grained versions of  static {\em clustering}. We explain the reasons behind this by performing a deeper analysis of the scheduling decisions taken for a DAG with $H=16$ and $\beta=512$ using the Gantt charts for {\em eager} {\em heft} and {\em cluster} scheduling depicted in	Fig. \ref{fig:combined_gannt}. The primary reason for the poor performance of  {\em eager} maybe attributed to the greedy selection of devices based on runtime availability rather than kernel preference. As a consequence, one can observe from Fig. \ref{fig:combined_gannt} (a), that multiple GEMM kernels have been scheduled on the CPU, thereby taking a significantly larger amount of time. Even though GPU bound GEMM kernels take less time, the execution of the callbacks are delayed since the CPU is being heavily used for the GEMM computation. This is evident from the gaps between the kernels scheduled on the GPU device. Since the callback function is initiated using a separate thread for notifying the host, it might happen that either i) the master thread running the $schedule$ routine is swapped out of main memory at that point of time or ii) there are not enough resources to spawn the thread for running the callback function. As a result, the read callbacks  wait before updating  $\mathcal{F}$ and $\mathcal{A}$ to allow progress. In contrast, we observe there are little to no gaps between kernels scheduled on the CPU device. This indicates that once  the \textit{ndrange} callback for a kernel executing on the CPU device finishes, it immediately updates $\mathcal{A}$. Consequently, the scheduling algorithm continues to dispatch kernels to the available CPU, thus causing starvation of the GPU resource.
	\par In contrast to {\em eager} scheduling, it may be observed from Fig. \ref{fig:combined_gannt}, that {\em heft} exclusively uses the GPU for the GEMM kernels and is thus approximately $2.4 \times$ faster than eager. But, {\em heft} fundamentally being a dynamic coarse-grained scheduling policy like {\em eager} still relies on read callbacks for dispatch decisions for every kernel.  The successive gaps introduced between each kernel execution in the DAG results in a considerable slowdown. 

	\par For our {\em clustering} scheme  we may observe from Fig. \ref{fig:combined_gannt} (c), that kernels start executing much later when compared to kernels being scheduled in the other schemes. This may be attributed to the fact, that our framework sets up the command queues first with operations pertaining to all kernels in a task component before actually dispatching the kernels to their respective devices. Another interesting observation in Fig. \ref{fig:combined_gannt} (c) is as follows. Since, the task component partitioning $\mathcal{T}$ ensures that there exists no inter edge buffers in $END(T)$ for each task component $T$ in the {\em clustering} scheme, there is no explicit requirement of callbacks which was the primary bottleneck in the other dynamic schemes. As a result, there exists no gaps between the execution of any two successive kernels in the DAG in Fig.\ref{fig:combined_gannt}. This holds true for the default coarse-grained configuration of {\em clustering} as well, since the task component partitioning set $\mathcal{T}$ is same as that of the fine-grained configuration. The definition of $\mathcal{T}$ coupled with intelligent use of command queues by the framework for {\em clustering} helps in avoiding these runtime delays and results in a considerable speedup when compared to dynamic scheduling decisions employed by {\em eager} and {\em heft}. The framework has been open-sourced in Github \footnote{{\tt https://github.com/anighose25/pyschedcl-concurrent}} with scripts for running all the experiments.

  %by having i) a static frontend that allows users to simply design specification files rather than complex host code for specifying the execution of a DAG and ii) a robust backend capable of taking intricate runtime decisions focused on extracting application level parallelism from the platform. %must be envisaged that take into account i)fine-grained scheduling decisions with respect to the structure of the DAG as well as ii) intricate runtime environment delays occurring as a result of enqueuing commands and processing simultaneous callback events. 
\section{Related Work} \label{sec:rel}%1.5-2 pages
	\par Given the rich API support of CUDA and OpenCL, several frameworks have emerged over the last few years with the objective of providing user friendly solutions for development of data parallel applications. Frameworks based on CUDA include OpenACC \cite{openacc} and HiCUDA \cite{hicuda} which supports a directive based programming model where relevant annotations in sequential C programs generate data parallel CUDA code for execution on the GPU. Frameworks such as GMAC \cite{gelado} ease programming by not requiring explicit memory operations while designing CUDA applications.   Several OpenCL based frameworks have also been envisioned in the past decade for general purpose heterogeneous programming. The most notable framework in this regard  is SkelCL \cite{skelCL} which offers support for designing algorithmic skeletons (higher order functions such as map, reduce, scan etc) which can be leveraged for implementing data parallel kernels. Note, the primary approach of this work is complementary in the sense that they focus on rapid kernel development, while our work focuses on scheduling optimizations on target heterogeneous architectures. The most recent work reported in \cite{pekka} proposes a novel set of APIs for specifying dependencies of a  DAG thus easing application development, but does not have explicit algorithm support for scheduling. The VirtCL framework \cite{virtcl} provides an abstraction layer between the programmer and the OpenCL runtime system acting as a hypervisor for scheduling multiple OpenCL applications. The abstraction framework leverages a profile driven history based scheduling scheme for dispatching OpenCL kernels on multiple devices. However, a major limitation for VirtCL is that it cannot operate with devices belonging to different platforms. Our framework in contrast is suited to work with different OpenCL platforms and supports both static and dynamic scheduling approaches for mapping OpenCL kernels. There also exists frameworks such as SnUCL \cite{snucl}, VOCL \cite{vocl}, MultiCL \cite{multicl} etc. that extend upon the OpenCL runtime API which  allows OpenCL applications to leverage devices belonging to heterogeneous clusters. While SnUCL and VOCL have no explicit algorithm support for scheduling, MultiCL relies on coarse-grained scheduling decisions with a focus on data partitioning across multiple devices.  Also, since these APIs are extensions of OpenCL, one cannot bypass the requirement of complex host program development for implementing data parallel applications.
	\par StarPU \cite{augonnet2011starpu, hugo2014composing} is a unified scheduling framework allowing users to design and experiment scheduling policies for both CUDA and OpenCL applications. The work reported in \cite{henry2014toward} presents an unified OpenCL implementation called SOCL which directly extends StarPU for exclusively supporting execution of OpenCL workloads across multiple devices. Both SOCL and StarPU rely on prior profiling information for each task on each device for constructing a performance model to be used for scheduling decisions and have algorithm support for coarse-grained scheduling decisions. In contrast, to the best of our knowledge, our framework is possibly the first to present an automated mechanism for enforcing fine-grained scheduling decisions which are relatively more adept in exploiting concurrency in OpenCL applications.
	%\par Despite the vast number of frameworks available, our proposed framework relies on specification files using which programmers can bypass the overhead of implementing complex host programs and design data parallel applications with ease. We also present a novel software architecture that is capable of automatically extracting both application level and platform level concurrency. Currently the scheduling schemes lack support for performance models necessary for obtaining near-optimal schedules. Future work entails investigating machine learning assisted techniques \cite{grewe2011static,grewe2013opencl,kofler2013automatic,smart,schedcl} for the same. We believe the robust API support in our framework  would allow researchers to investigate these avenues and accordingly design and validate novel scheduling algorithms for heterogeneous platforms.
	\section{Conclusion} \label{sec:conclusion}
	We propose a platform agnostic scheduling framework that not only enables users to design HPC applications with ease, but also performs optimized scheduling decisions that exploit both application-level and platform-level concurrency. For an application with ample scope for concurrency, we have observed that rather than relying on traditional coarse-grained  scheduling decisions, implementing fine-grained scheduling policies using {\em PySchedCL} where the user specifies an intuitive task component partitioning $\mathcal{T}$ after examining the structure of a DAG application results in significantly better execution times. Future work entails investigating sophisticated low-level scheduling approaches such as sub-kernel partitioning \cite{ccuda,ktile} at the work-item level  for effective interleaving of concurrent kernels. Such approaches coupled with Machine Learning assisted control theoretic scheduling solutions \cite{caloree} shall be used to develop an auto-tuning framework on top of {\em PySchedCL} which would automatically determine given an application-architecture pair, the optimal allocation of command queues across devices in the platform.
	% that take into account the impact of memory and register usage while  interleaving kernels concurrently on the same GPU. These low-level schemes coupled with Control-theoretic and Machine Learning based modelling techniques shall be used to develop an autotuning framework on top of {\em PySchedCL}. The final objective would be to automatically determine given an application-architecture pair, which scheduling policy should be followed and what number of command queues per device should be assigned in order to maximally exploit the underlying hardware for executing a data-parallel application efficiently. 
	
	\bibliographystyle{ieeetr}
	\bibliography{main.bib}
	\vspace{-1.4cm}

\end{document}